\documentclass{iopart}
\usepackage{iopams, setstack}
\usepackage{bm,hyperref}
\usepackage{graphicx,subfigure,wrapfig}

\newcommand{\eqref}[1]{{(\ref{#1})}}    

\newcommand{\rmS}{{\rm S}}

\newcommand{\rmR}{{\rm R}}
\newcommand{\rmD}{{\rm D}}

\newcommand{\ItP}{{P}} 
\newcommand{\ItF}{{\mathcal{F}}} 
\newcommand{\ItE}{{\mathcal{E}}}

\newcommand{\Lie}{{\mathcal{L}}} 
\newcommand{\LieX}{{\mathcal{L}_{\xi}}} 

\begin{document}
\title{Mechanics of extended masses in general relativity}
\author{Abraham I. Harte}
\address{Max-Planck-Institut f\"{u}r Gravitationsphysik, Albert-Einstein-Institut Potsdam-Golm, Germany}
\ead{harte@aei.mpg.de}

\begin{abstract}
The ``external'' or ``bulk'' motion of extended bodies is studied in general relativity. Compact material objects of essentially arbitrary shape, spin, internal composition, and velocity are allowed as long as there is no direct (non-gravitational) contact with other sources of stress-energy. Physically reasonable linear and angular momenta are proposed for such bodies and exact equations describing their evolution are derived. Changes in the momenta depend on a  certain ``effective metric'' that is closely related to a non-perturbative generalization of the Detweiler-Whiting R-field originally introduced in the self-force literature. If the effective metric inside a \textit{self-gravitating} body can be adequately approximated by an appropriate power series, the instantaneous gravitational force and torque exerted on it is shown to be identical to the force and torque exerted on an appropriate \textit{test} body moving in the effective metric. This result holds to all multipole orders. The only instantaneous effect of a body's self-field is to finitely renormalize the ``bare'' multipole moments of its stress-energy tensor. The MiSaTaQuWa expression for the gravitational self-force is recovered as a simple application. A gravitational self-torque is obtained as well. Lastly, it is shown that the effective metric in which objects appear to move is approximately a solution to the vacuum Einstein equation if the physical metric is an approximate solution to Einstein's equation linearized about a vacuum background.
\end{abstract}

\pacs{04.20.Cv, 04.25.-g, 04.40-b, 45.20.-d}

\vskip 2pc

\section{Introduction}

Newtonian celestial mechanics typically describes the motion of widely-separated masses using two types of parameters (see, e.g., \cite{Damour300, Dix79}). These concern either the behavior of each body as a whole -- the ``external'' or ``bulk'' parameters -- or the details of their internal dynamics. Examples of external parameters are the center of mass positions and spin angular momenta of the various masses. The internal variables include, e.g., the density and velocity distributions inside each body. In typical applications, there is very little coupling between the internal and external parameters. As a consequence, one can often compute the center of mass positions of each extended body in an $N$-body system as though that system were composed of point particles described only by their positions and masses. Furthermore, the spin angular momentum of each body in such a system can usually be taken to remain constant (and does not affect the center of mass motion). Both of these statements are, of course, approximate. A more accurate description requires introducing additional parameters such as quadrupole moments. These depend on the internal dynamics, but in a relatively mild way that often lends itself to simple phenomenological models.

The external variables decouple from the internal ones in Newtonian gravity largely because the net force and torque exerted by a body's self-field always vanishes. There are no self-forces or self-torques in this theory\footnote{The self-force is defined here as the net force exerted by the self-field. The self-field is, in turn, defined in a standard way. See Sect. \ref{Sect:NewtGrav} below. Note that this definition of self-force is not the same as the perturbative one used in, e.g., \cite{NewtSF}.}. The instantaneous evolution equations for an object's linear momentum, center of mass position, and spin do not explicitly involve its self-field. These quantities \textit{are} affected by the self-field through its action on other objects, although this is an effect that takes time to accumulate. 

\textit{A priori}, it is not clear that similar statements can be made for matter interacting with relativistic fields. Such fields carry energy and momentum, so self-forces arise generically. This does not, however, preclude an internal-external  split of the dynamics. The usefulness of such a split does not require that self-forces vanish entirely, but only that they do not depend in any essential way on the details of a body's internal structure.

To illustrate this point, consider the motion of a small electric charge in approximate internal equilibrium moving non-relativistically in flat spacetime. It has long been known that under suitable conditions, the center of mass acceleration $\mathbf{a}(s)$ of such a charge at time $s$ very nearly satisfies\footnote{This equation has been established as a valid approximation only for the \textit{acceleration} of a physical charge (see, e.g., \cite{GrallaHarteWald, HarteEMNew}). This does not mean that a trajectory with an acceleration satisfying \eqref{ALD} for all time is guaranteed to stay near the physical trajectory. Many such motions violate the conditions under which the equation was derived (even on short timescales), and must therefore be discarded. Additionally, there may be neglected terms which lead to qualitatively different behavior over long times. Better-behaved equations arise by ``reducing order'' \cite{GrallaHarteWald, Spohn}, which changes \eqref{ALD} only by an amount comparable to the error terms that are already present. This leads to an equation often attributed to Landau and Lifshitz in the relativistic case \cite{LL}.}
\begin{equation}
	m \mathbf{a} = \mathbf{F}_{\mathrm{ext}} + \frac{2}{3} q^{2} \frac{\rmd \mathbf{a}}{\rmd s}  - \delta m \mathbf{a}. \label{ALD}
\end{equation} 
Here, $\mathbf{F}_{\mathrm{ext}}$ is an externally-imposed force, $q$ is the object's total charge and $m$ its (bare) mass. The last two terms on  the right-hand side of this equation arise from interactions with the body's own electromagnetic field. The first of these is ``simple'' in that it depends only on bulk parameters -- namely $q$ and $\mathbf{a}$ -- already required to describe the motion of a charged test particle.

$\delta m$, by contrast, has a very different character. In an appropriate approximation, it is the self-energy of the charge distribution as it would typically be defined \cite{HarteEMNew}. Denoting the electric charge density by $\rho_e$,
\begin{equation}
	\delta m (s)= \frac{1}{2} \int \rmd^3 \mathbf{X} \int \rmd^3 \mathbf{X}' \left( \frac{ \rho_e (\mathbf{X}, s) \rho_e (\mathbf{X}', s) }{ | \mathbf{X} - \mathbf{X}' | } \right) .
\end{equation}
It is clear that $\delta m$ depends on the body's internal structure in a nontrivial way. Despite this, \eqref{ALD} may be rewritten in the form
\begin{equation}
	\hat{m} \mathbf{a} = \mathbf{F}_{\mathrm{ext}} + \frac{2}{3} q^{2} \frac{\rmd \mathbf{a}}{\rmd s}, 
\end{equation}
where $\hat{m} := m + \delta m$ is interpreted as a renormalized or effective mass. The same assumptions leading to the derivation of \eqref{ALD} can also be used to show that $\rmd \hat{m}/\rmd s = 0$. For a well-behaved extended object,  $\delta m$ is finite. The mass renormalization effect is therefore finite as well.

Even though the self-force is significant in this example (and depends on nontrivial details of the body's internal structure), the final equation of motion involves only parameters of the same type as those already needed to describe the motion of a charged test particle. To the extent that \eqref{ALD} can be trusted, this means that the external variables largely decouple from the internal ones in electromagnetism. The center of mass acceleration of an appropriate extended self-interacting charge distribution is the same as the acceleration of a monopole test charge moving in an effective electric field given by the external one plus $\frac{2}{3} q \rmd \mathbf{a}/\rmd s$ (at the particle's location). This effective field may be shown to arise naturally as a certain solution to the vacuum Maxwell equations \cite{HarteEMNew, Dirac, DetWhiting}.

This result can be generalized considerably. Essentially all restrictions regarding the charge's size, internal dynamics, and speed may be removed. For almost any bounded \textit{self-interacting} charge-current distribution in flat spacetime, physically reasonable linear and angular momenta may be defined that evolve as though they were the momenta of an extended \textit{test} charge (or a pointlike test charge ``with structure'') moving in a certain effective electromagnetic field \cite{HarteEMNew}. This effective field satisfies the vacuum Maxwell equations near the charge. All effects of the self-force and self-torque can be non-perturbatively absorbed into the definitions of the momenta and the effective field. Whether or not the internal structure is ``effaced'' from the external laws of motion therefore reduces to a question regarding the nature of the effective field. In all but the most extreme systems, the effective field may be shown to depend only on bulk parameters like the total charge. Very similar results also hold in generic (but fixed) curved spacetimes. The only qualitative change that occurs when introducing spacetime curvature is that the quadrupole and higher multipole of a charge's stress-energy tensor are renormalized along with its momenta \cite{HarteHighMult}. Analogous statements are known for matter interacting with linear scalar fields as well \cite{HarteHighMult, HarteScalar}.

Results of this type greatly expand the scope of -- and provide a basis for -- what has been referred to as the Detweiler-Whiting axiom \cite{DetWhiting, PoissonRev}. It is well-known that point particles are incompatible with, e.g., the standard formulation of Maxwell electrodynamics (and with general relativity \cite{GerochTraschen}). Despite this, ``point particle methods'' can still be used if additional axioms are introduced into the theory. Suppose, for example, that a certain portion of the self-field associated with a pointlike electric charge is assumed not to affect its motion. Detweiler and Whiting considered this possibility with an ignorable field constructed using a certain symmetric Green function \cite{DetWhiting}. Subtracting this field from the physical one leaves a result which is easily calculated and well-behaved. It also satisfies the vacuum Maxwell equations at the location of the particle. Substituting this difference field into the Lorentz force equation produces the standard Dewitt-Brehme result \cite{PoissonRev, DewittBrehme} for the motion of a self-interacting charged particle in curved spacetime. Similar subtractions were also used to efficiently reproduce equations of motion that had previously been derived for self-interacting scalar charges as well as uncharged masses in linearized general relativity. 

The results of \cite{HarteEMNew, HarteScalar} show that this ability to ignore what is referred to as the Detweiler-Whiting S-field is not merely a computational shortcut allowing the use of point particle methods in cases where actual point particles cannot exist. A very general type of ``Detweiler-Whiting axiom'' may be rigorously \textit{derived} from first principles for a large class of extended scalar and electromagnetic charge distributions moving in fixed spacetimes. This paper uses similar methods to treat the gravitational problem. Specifically, it investigates whether the bulk dynamics of an uncharged mass in general relativity can be reduced to test body motion in an effective metric (in a nontrivial sense\footnote{Suppose that it is known, for example, that the acceleration of an extended nonrelativistic electric charge satisfies $m \mathbf{a}  = q \mathbf{E} + \mathbf{f}$. If $q \neq 0$, this is trivially equivalent to the motion of a pointlike test body in the field $\mathbf{E} + q^{-1} \mathbf{f}$. In general relativity, equations with the form $m \rmD u^a(s)/\rmd s = f^a$ can always be rewritten as geodesic equations associated with some connection. It is only in special cases, however, that such identifications are useful.}). Related questions have been studied in various contexts using the post-Newtonian approximation \cite{Damour300, LeviCivita1PN, Damour83, WillEquiv}, where they are often referred to as ``effacement principles'' or demonstrations of the strong equivalence principle. 

The work presented here is motivated more by the types of systems commonly encountered in discussions of the gravitational self-force. These discussions typically allow the body of interest to move at relativistic speeds in a strongly curved background spacetime, but restrict it to be small compared to all scales associated with that background. One also assumes that the internal structure of the body does not vary too rapidly. Under these conditions -- made precise in, e.g., \cite{GrallaWald, PoundSF} -- an equation of motion may be derived that does not depend on any details of the body's internal structure. At lowest order, it is just the geodesic equation associated with the background spacetime. The next approximation introduces forces due to both gravitational self-interaction and spin. The latter effect is the Papapetrou force long known to act on spinning test particles \cite{Mathisson, Papapetrou}. The self-force component is  typically referred to as the MiSaTaQuWa force after the authors who originally obtained it: Mino, Sasaki, Tanaka, Quinn, and Wald \cite{MST, QuinnWald}. Neglecting the Papapetrou term, the motion is most naturally viewed as geodesic with respect to a certain effective metric satisfying the linearized vacuum Einstein equation \cite{DetWhiting}.

We show that this is a special case of a much more general result. Certain definitions of linear and angular momentum are proposed for extended compact matter distributions in general relativity. It is assumed that there is no stress-energy near the object of interest other than its own (except perhaps dark energy equivalent to a cosmological constant). An effective metric $\hat{g}_{ab}$ is then defined based on a non-perturbative generalization of the Detweiler-Whiting decomposition of the physical metric $g_{ab}$. There is a sense in which the force and torque depend only on $\hat{g}_{ab}$ and the details of the body's stress-energy tensor. If $\hat{g}_{ab}$ varies sufficiently slowly that it can be expanded in a Taylor series about an appropriate point inside the body (in a Riemann normal coordinate system constructed using $\hat{g}_{ab}$), the instantaneous force and torque are shown to be identical to those of an appropriate test body moving in the effective metric. A similar result also holds for a certain definition of the center of mass. This means that equations known to hold for test bodies (possibly with higher multipole moments) also hold for masses with significant self-interaction.

As a simple application, note that the simplest test bodies move on geodesics. The simplest self-interacting bodies therefore move on geodesics of the effective metric. The MiSaTaQuWa expression for the gravitational self-force follows easily with some minor additional assumptions. Similarly, the simplest equations for a body's spin evolution are those of parallel transport. Appropriate self-gravitating masses therefore parallel-propagate their spins in the effective metric. Corrections to these statements arising from higher order multipole moments are easily added when appropriate.

The assumptions adopted here are different from those found in other treatments of the gravitational self-force. Most importantly, the approaches of, e.g., \cite{GrallaWald, PoundSF} are intrinsically perturbative. They work in an intermediate ``buffer'' region outside of the body of interest and assume that the metric there is a small perturbation off of some vacuum background. Such methods can be applied even to the motion of black holes, which lies beyond the scope of the formalism developed here. We require that a body be described by a well-behaved stress-energy tensor. Despite this restriction, there are considerable benefits to our assumptions. They allow the analysis of objects that may be highly distorted and dynamical: Dixon's multipole expansions \cite{Dix79, Dix74} for the motion of extended test masses are generalized to all orders. Explicit formulae for the momenta are also provided in terms of the body's internal structure. There are aesthetic advantages as well. In regimes where they overlap, the method presented here requires far less computation than others in the literature. It also provides significantly more physical insight.

\subsection*{Layout of the paper}

The main results established in this work are obtained using only a modicum of computation. Despite this, a number of concepts and techniques are employed that are not in common use. While not new \cite{HarteEMNew, HarteHighMult, HarteScalar, HarteSyms}, the relevant ideas are reviewed in Sect. \ref{Sect:TestMass} by applying them to problems with which the reader might be more familiar. Sect. \ref{Sect:NewtGrav} starts by discussing in detail the motion of self-interacting masses in Newtonian gravity. While the conclusions of this section are completely standard, the formulation used to obtain them has several unusual features. It does not, for example, rely on any choice of coordinates. It also treats a body's linear and angular momenta as different components of a single scalar functional on the space of Euclidean Killing fields. The lack of self-forces and self-torques is shown to follow from the symmetries of a particular Green function used to define what is meant by the term ``self-field.'' 

Similar techniques are used in Sect. \ref{Sect:TestGR} to discuss the motion of fully relativistic extended \textit{test} masses in curved spacetimes. This is a review of appropriate aspects of Dixon's work on the subject \cite{Dix79, Dix74, Dix70a} as reformulated in \cite{HarteSyms}. Linear and angular momenta are shown to arise as two components of a scalar functional that now takes as input certain ``generalized Killing fields'' when ordinary Killing fields do not exist. These vector fields are defined in detail in the appendix. Using them, multipole expansions for the force and torque are established when appropriate. A center of mass is also defined, and it is pointed out that the hidden momentum is generically nonzero (i.e., the linear momentum is not parallel to the center of mass velocity). Familiarity with the discussion of Sect. \ref{Sect:TestMass} is essential for understanding the remainder of this paper, where similar techniques are used to analyze the motion of self-gravitating masses in general relativity. 

The main results of this paper are contained in Sect. \ref{Sect:MechanicsGen}. After discussing the problem of defining an effective metric and a self-field abstractly in Sect. \ref{Sect:InitConsider}, a specific definition for the self-field is given in Sect. \ref{Sect:SFDef}. Exact expressions for the force and torque are then derived in Sect. \ref{Sect:FreeFall}. Multipole expansions of these equations are performed in Sect. \ref{Sect:Multipole}. A brief discussion of the multipole moments appearing in the resulting series is given in Sect. \ref{Sect:EffMomenta}. Finally, a center of mass is defined in Sect. \ref{Sect:CM} and related to the linear momentum.

Sect. \ref{Sect:Dipole} applies these results to the motion of a small body. The monopole-dipole approximation is discussed in various ways. We then specialize to general relativity linearized off of a vacuum background and derive an equation for the center of mass position that includes the MiSaTaQuWa ``self-force.'' A similar result is also obtained for the spin evolution (including a ``self-torque'').

The sign conventions used here are those of Wald \cite{Wald}. Metrics therefore have signature $+2$ and the Riemann tensor satisfies $2 \nabla_{[a} \nabla_{b]} \omega_{c} = R_{abc}{}^{d} \omega_{d}$ for any 1-form $\omega_{a}$. The Ricci tensor is given by $R_{ab} = R_{acb}{}^{c}$. Multiple metrics are discussed in this paper, so indices are not raised and lowered unless indicated otherwise. In almost all cases, factors of the appropriate metric are displayed explicitly. There are three main metrics that appear: $g_{ab}$ denotes the full physical metric, $\hat{g}_{ab}$ a certain effective metric, and $\bar{g}_{ab}$ a background metric. Derivative operators and curvature tensors associated with the latter two geometries are distinguished with a hat or bar as appropriate. Non-geometric quantities (like momenta) with hats typically denote renormalized or effective versions of their plainer counterparts. Abstract indices are written using letters from beginning of the Latin alphabet, while Greek indices represent spacetime coordinate components. The letters $i,j, \ldots$ represent spatial coordinate components. Units are used where $G=c=1$.

\section{Motion in simple cases} 
\label{Sect:TestMass}

The main goal of this paper is to describe, in some sense, the large-scale or bulk motion of extended masses in general relativity. This is done by analyzing quantities that may be interpreted as a body's net linear and angular momenta (as well as the closely related notion of its center of mass). 

The type of momentum considered here is similar to the one developed by Dixon \cite{Dix79, Dix74, Dix70a}. Mathematically, Dixon's momenta are tensor fields defined non-perturbatively along a preferred worldline in the physical spacetime. They take as input this worldline and a timelike vector field prescribed along it. The linear or angular momentum of an extended body is then computed by integrating its stress-energy tensor over a spacelike hypersurface in a particular way. The evolution of these quantities is strongly constrained by stress-energy conservation.

The only significant restriction to the use of Dixon's momenta is that an object's stress-energy tensor be bounded in spatial directions. This bound is also required not to be ``extremely large''  in a particular sense \cite{Dix74, CM}. Limitations on the metric are minimal. Despite this, most applications (e.g., \cite{HarteQuadrupole, ItalianQuadrupole}) have been restricted to the test body regime where the body of interest is not allowed to backreact onto the geometry. While Dixon's momenta retain a number of interesting properties in a more a general context \cite{Dix79, Dix74, SchattnerStreubel1,SchattnerStreubel2}, other characteristics are less satisfactory. For example, it has been shown that even in flat spacetime electromagnetism, the momenta do not behave as simply as might have been expected once electromagnetic self-interaction is taken into account \cite{HarteEMOld}. This problem can be eliminated with a relatively simple modification \cite{HarteEMNew}.

Similar changes are proposed here in order to obtain physically reasonable momenta that obey simple evolution equations in the presence of significant gravitational self-interaction (but without electromagnetic or other long-range non-gravitational fields). The basic strategy is to first postulate ``bare'' momenta. These agree with Dixon's definitions in the test mass regime, but differ in general. The important point is that the evolution equations for the bare momenta include total time derivatives of certain terms involving parts of the self-field. These derivatives are easily eliminated by redefining the momenta. The resulting variables obey simple evolution equations in a wide variety of contexts.
 
This section sets the foundation for deriving these results by reviewing the relevant techniques in simpler systems. We start by discussing the motion of self-interacting extended masses in Newtonian gravity. This is carried out from a somewhat unusual point of view introduced in \cite{HarteScalar}. Similar techniques are then used to analyze relativistic motion in a curved spacetime, but without self-interaction. The resulting definitions and conclusions are equivalent to Dixon's \cite{Dix79, Dix74, Dix70a}. Many aspects of the formalisms discussed in this section carry through almost without modification to cases involving self-interaction in general relativity.

\subsection{Self-interacting masses in Newtonian gravity}
\label{Sect:NewtGrav}

As a first step to understanding the motion of extended bodies in general relativity, consider the motion of a freely-falling extended mass in Newtonian gravity. Such a mass may be modelled as having a time-dependent configuration on a three-dimensional Euclidean space $(\mathcal{M}, g_{ab})$. We assume that this configuration has nonzero volume and may be entirely contained in a compact region $\Sigma_s \subset \mathcal{M}$ at time $s$ (i.e., the body is extended but with a finite size). Also suppose that the body of interest is composed of a single material with mass density $\rho(x,s)$ and 3-velocity $v^a(x,s)$. Given any $x \in \mathcal{M}$ and $s \in \mathbb{R}$, $\rho(x,s) \geq 0$. While it is possible to relax this requirement considerably, we assume for simplicity that $\rho(x,s)$ and $v^a(x,s)$ are smooth in both of their arguments. Lastly, suppose that there exists an open neighborhood of $\Sigma_s$ containing no matter other than the body of interest. This ensures that there is no direct contact with other objects.

In general, the density and velocity distributions are constrained by local mass and momentum conservation. These laws have the explicit forms
\begin{equation}
	\frac{\partial \rho}{\partial s} + \nabla_a (\rho v^a) = 0,
	\label{MassCons}
\end{equation}
and 
\begin{equation}
	\frac{ \partial }{ \partial s} (\rho v^a) + \nabla_b ( \rho v^a v^b  + \Sigma^{ab} ) = - \rho g^{ab} \nabla_b \phi, 
	\label{MomentumCons}
\end{equation}
where $\nabla_a$ denotes the Levi-Civita connection associated with the Euclidean 3-metric $g_{ab}$. $\Sigma^{ab} = \Sigma^{(ab)}(x,s)$ represents the body's stress tensor, $\phi(x,s)$ the gravitational potential, and $g^{ab}(x)$ the inverse of $g_{ab}(x)$.  Besides these equations, the body of interest is also assumed to be a source for the gravitational field. The potential therefore satisfies 
\begin{equation}
	g^{ab} \nabla_a \nabla_b \phi = 4 \pi \rho
	\label{Poisson}
\end{equation}
throughout $\Sigma_s$.

The simplest consequence of these equations is that the total mass $m$ cannot change. Using $\rmd V$ to denote the natural (three-dimensional) volume element associated with $g_{ab}$, let
\begin{equation}
	m := \int_{\Sigma_s} \rho(x,s) \rmd V.
\end{equation}
It immediately follows from \eqref{MassCons} that $m$ is independent of $s$.

Eq. \eqref{MomentumCons} constrains the evolution of the body's total linear momentum $p^a$ and angular momentum $S_{a}$. The linear momentum is typically defined by integrating the components $\rho v^i$ in a Cartesian coordinate system $X^i(x)$. A similar integral also exists for the angular momentum. It is important for later generalizations to avoid any coordinate choices such as these and instead define the momenta geometrically. This can be accomplished by recalling that global linear momentum conservation is associated with the translational invariance of Euclidean space. Similarly, global angular momentum conservation is related to the rotational invariance of Euclidean space. Translations and rotations together comprise the continuous isometries of $(\mathcal{M}, g_{ab})$. Generators of these isometries are Killing vectors.

Given any Euclidean Killing field $\xi^a$ and time $s$, consider
\begin{equation}
	P_\xi (s) := \int_{\Sigma_s} \rho(x,s) v^a(x,s) g_{ab}(x) \xi^b(x) \rmd V. 
	\label{PXiNewt}
\end{equation}
This is a linear functional on the six-dimensional space of Killing fields. It may be viewed as returning the component of momentum ``conjugate'' to $\xi^a$. If, say, $\zeta^a = \partial/ \partial X^1$ is a particular Killing field associated with translations in the $X^1$-direction, $P_{\zeta}(s)$ is equal to the Euclidean component $p_1(s) = g_{ab} p^a \zeta^b$ of the body's linear momentum as it would ordinarily be defined. Similarly, use of a purely rotational Killing field in \eqref{PXiNewt} returns a component of the body's angular momentum. 

In general, $P_\xi$ is equal to a sum of linear and angular momentum components. The precise form of this sum may be established by recalling that any Killing field is fixed everywhere by specifying it and its first derivative at a single point \cite{Wald}. Choosing a (possibly time-dependent) origin $\gamma_s \in \mathcal{M}$,
\begin{equation}
	\xi^a (x) \Leftrightarrow \{ \xi^a(\gamma_s) , \nabla_b \xi^a (\gamma_s) \}.
	\label{KillingDataEucl}
\end{equation}
This correspondence may be observed explicitly by noting that in Cartesian coordinates $X^i(x)$, all Euclidean Killing fields have the form
\begin{equation}
	\xi^i (x) = \xi^i (\gamma_s) + [X^j(x) - X^j(\gamma_s)] \partial_j \xi^i (\gamma_s)  .
\label{EuclKilling} 
\end{equation}
$\xi^i (\gamma_s)$ may be chosen arbitrarily in this equation, while $\partial_j \xi^i (\gamma_s)$ must be an antisymmetric matrix. Note that $\xi^i(x)$ is linear in the ``data'' $\{ \xi^i(\gamma_s), \partial_j \xi^i(\gamma_s)\}$. This is a generic feature of the correspondence \eqref{KillingDataEucl}, and is unrelated to working in Euclidean space. 

Now, it is clear from \eqref{PXiNewt} that $P_\xi$ is linear in $\xi^a(x)$. It follows that $P_\xi$ may always be written as a linear combination of $\xi^a(\gamma_s)$ and $\nabla_b \xi^a (\gamma_s)$. The appropriate coefficients are essentially the linear and angular momenta as they would typically be defined. If $\Xi_a := g_{ab} \xi^b$, let $p^a$ and $S^{ab} = S^{[ab]}$ satisfy
\begin{equation}
	P_\xi(s) = p^a(\gamma_s,s) \Xi_a (\gamma_s) + \frac{1}{2}  S^{ab} (\gamma_s, s) \nabla_{a}  \Xi_{b}(\gamma_s).
\label{PtopSNewt}
\end{equation}
All Killing fields may be generated by varying $\Xi_a(\gamma_s)$ amongst all possible 1-forms and $\nabla_a \Xi_b = \nabla_{[a} \Xi_{b]} (\gamma_s)$ amongst all possible 2-forms. Knowledge of $P_\xi$ for all $\xi^a$ is therefore equivalent to knowledge of $p^a$ and $S^{ab}$. The angular momentum 1-form $S_a$ may be extracted from $S^{ab}$ via
 \begin{equation}
 	S_a := \frac{1}{2} \epsilon_{abc} S^{bc},
 \label{StoSTensNewt}
 \end{equation}
where $\epsilon_{abc}$ denotes the natural volume element associated with $g_{ab}$. Both $S_a$ and $S^{ab}$ contain all angular momentum information in the three dimensions considered here. Eq. \eqref{StoSTensNewt} may therefore be inverted, giving $S^{ab} = \epsilon^{abc} S_c$. 

Eqs. \eqref{PXiNewt} and \eqref{PtopSNewt} provide a coordinate-invariant definition for a body's linear and angular momenta. Mathematically, $p^a(\gamma_s,s)$ and $S^{ab}(\gamma_s,s)$ are tensors at the point $\gamma_s$. In more elementary presentations of Newtonian mechanics, $\gamma_s$ corresponds to the preferred point required to define the angular momentum. It is often taken to coincide with the object's center of mass, although this choice is not essential.

Later sections in this paper makes extensive use of functionals like $P_\xi$. For a relativistic object moving in curved spacetime, the quantity $\rho v^a$ appearing in the integrand of \eqref{PXiNewt} is translated into an obvious analog involving the body's stress-energy tensor. More difficult to generalize is $\xi^a$, which must be chosen from a suitable space of ``generalized Killing fields'' when ordinary Killing fields do not exist. This is, however, a surmountable problem. The relation \eqref{PtopSNewt} between $P_\xi$, $p^a$, and $S^{ab}$ does not change at all for relativistic motion in curved spacetimes.

Even in the Newtonian context, there are advantages to working with $P_\xi$ rather than $p^a$ and $S^{ab}$. Most importantly, this functional allows the linear and angular momenta to be manipulated simultaneously merely by performing operations on scalars. $P_\xi$ also provides a clear relation between symmetries and conservation laws.

It is now possible to discuss how a Newtonian body's momenta vary over time. These changes can be extracted from changes in $P_\xi$. Using \eqref{PXiNewt} together with \eqref{MomentumCons} shows that
\begin{equation}
	\frac{\rmd}{\rmd s} P_\xi(s) = - \int_{\Sigma_s} \rho(x,s) \LieX \phi(x,s) \rmd V,
	\label{PDotNewt}
\end{equation}
where $\LieX \phi$ denotes the Lie derivative of $\phi$ with respect to $\xi^a$. Differentiating \eqref{PtopSNewt} also shows that
\begin{equation}
	\frac{\rmd}{\rmd s} P_\xi(s) = \frac{\rmD p^a}{\rmd s} \Xi_a + \frac{1}{2} \left( \frac{\rmD S^{ab}}{\rmd s} - 2 p^{[a} \dot{\gamma}_s^{b]} \right) \nabla_a \Xi_b,
	\label{PDotFNNewt}
\end{equation}
where $\dot{\gamma}_s^a := \rmd \gamma_s^a/\rmd s$ and we have used the fact that second derivatives of Killing fields vanish in flat space. Equating the right-hand side of this equation with the right-hand side of \eqref{PDotNewt} produces evolution equations for both $p^a$ and $S^{ab}$. It is useful to define a force $F^a$ and torque $N^{ab} = N^{[ab]}$ such that
\begin{equation}
	\frac{\rmd}{\rmd s} P_\xi(s) = F^a \Xi_a + \frac{1}{2} N^{ab} \nabla_a \Xi_b,
	\label{PtoForceNewt}
\end{equation}
Then
\begin{eqnarray}
	\frac{ \rmD p^a }{\rmd s } = F^a,
	\\
	\frac{\rmD S^{ab} }{\rmd s} = 2 p^{[a} \dot{\gamma}^{b]}_s + N^{ab},
\end{eqnarray}
where $\rmD/\rmd s$ denotes the covariant path derivative associated with $g_{ab}$. Just as knowledge of $P_\xi$ is equivalent to knowledge of $p^a$ and $S^{ab}$, \eqref{PtoForceNewt} provides a one-to-one correspondence between $\rmd P_\xi /\rmd s$ and $F^a$ and $N^{ab}$. The (possibly unfamiliar) term involving $p^{[a} \dot{\gamma}^{b]}_s$ in the evolution equation for the angular momentum measures the degree to which $p^a$ and $\dot{\gamma}_s^a$ fail to be collinear. This term vanishes here if $\gamma_s$ is chosen to coincide with the body's center of mass. In the relativistic context, it rarely vanishes exactly.

Now note that the evolution equation \eqref{PDotNewt} for $P_\xi$ is linear in $\phi$. It therefore makes sense to discuss the force and torque exerted by particular components of the potential. Consider, in particular, the effect of the self-field $\phi_\rmS$. This is defined using a symmetric Green function $G_\rmS(x,x') = G_\rmS(x',x)$ that satisfies
\begin{equation}
	g^{ab} \nabla_a \nabla_b G_\rmS (x,x')=  - 4 \pi \delta(x,x')
	\label{BoxGNewt}
\end{equation}
and vanishes when its arguments are infinitely separated. In Cartesian coordinates $X^i(x)$, it is explicitly
\begin{equation}
	G_\rmS(x,x') = \frac{1}{|\mathbf{X}(x) - \mathbf{X}(x')|}.
\end{equation}
The self-field $\phi_\rmS$ is now defined by
\begin{equation}
	\phi_\rmS(x,s) := - \int_{\Sigma_s} \rho(x',s) G_\rmS (x,x') \rmd V'.
	\label{PhiS}
\end{equation}
It is clear from \eqref{BoxGNewt} that
\begin{equation}
	g^{ab} \nabla_a \nabla_b \phi_\rmS = 4 \pi \rho
\end{equation}
in $\Sigma_s$. Combining this with \eqref{Poisson} shows that the difference field
\begin{equation}
	\hat{\phi} := \phi - \phi_\rmS
	\label{PhiHat}
\end{equation} 
satisfies the vacuum equation 
\begin{equation}
	g^{ab} \nabla_a \nabla_b \hat{\phi} =0 
	\label{BoxPhiHat}
\end{equation}
in $\Sigma_s$.

Inserting \eqref{PhiS} and \eqref{PhiHat} into \eqref{PDotNewt} and commuting integrals shows that
\begin{equation}
	\frac{ \rmd P_\xi }{ \rmd s } = - \int_{\Sigma_s} \rmd V \rho \LieX \hat{\phi}- \frac{1}{2} \int_{\Sigma_s} \rmd V \int_{\Sigma_s} \rmd V' \rho \rho' \LieX G_\rmS ,
	\label{PDotNewt2}
\end{equation}
where $\rho' := \rho(x',s)$. This equation involves the Lie derivative $\LieX G_\rmS(x,x')$ of a two-point scalar $G_\rmS(x,x')$. Unless otherwise noted, Lie derivatives of objects depending on multiple points are defined in this paper to act on each of those points individually. For example,
\begin{equation}
	\LieX G_\rmS (x,x') = \xi^a (x) \nabla_a G_\rmS (x,x') + \xi^{a'} (x') \nabla_{a'} G_\rmS (x,x').
	\label{LieTwoPoint}
\end{equation}
The integrals involving this term in \eqref{PDotNewt2} determine the force and torque exerted by the self-field (i.e., the self-force and self-torque). These are very simple to evaluate. $G_\rmS$ is invariant with respect to all translations and rotations, so 
\begin{equation}
	\LieX G_\rmS = 0
	\label{LieGNewt}
\end{equation}
for all Killing fields $\xi^a$. All self-forces and self-torques therefore vanish. The momenta satisfy
\begin{equation}
	\frac{ \rmd }{ \rmd s } P_\xi = - \int_{\Sigma_s} \rmd V \rho \LieX \hat{\phi}. 
	\label{PDotNewt3}
\end{equation}
 Instantaneously, this is the same as the equation satisfied by a \textit{test} body with density $\rho$  immersed in the (vacuum) potential $\hat{\phi}$.

Elementary discussions of Newtonian gravity commonly ascribe vanishing self-forces and self-torques to Newton's third law. It is instructive to note that this concept is equivalent to the symmetry \eqref{LieGNewt} of the Green function used to define the self-field. To see this, consider two small volumes $\rmd V$ and $\rmd V'$. In Cartesian coordinates, the gravitational force exerted \textit{on} matter in $\rmd V$ \textit{by} matter in $\rmd V'$ is reasonably interpreted to refer to
\begin{equation}
	\rho(x,s) \rho(x',s) \partial_{i} G_\rmS (x,x') \rmd V \rmd V'.
	\label{Newton3Part1}
\end{equation}
Now consider only the first coordinate component of the force exerted by $\rmd V$ on $\rmd V'$. Adding to this the first coordinate component of the force exerted on $\rmd V$ by $\rmd V'$ results in
\begin{equation}
	\rho(x,s) \rho(x',s) \mathcal{L}_\zeta G_\rmS(x,x') \rmd V \rmd V' = 0,
\end{equation}
where $\zeta^a = \partial/\partial X^1$ is a translational Killing vector. This argument may obviously be repeated for any translational Killing field $\zeta^a$ (i.e., for any $\zeta^a$ satisfying $\nabla_b \zeta^a =0$). It follows that the force on $\rmd V$ due to $\rmd V'$ is equal and opposite to the force on $\rmd V'$ due to $\rmd V$. Considering translational Killing fields in \eqref{LieGNewt} therefore implies the weak form of Newton's third law. That $G_\rmS$ is also invariant under rotations implies the strong form of Newton's third law: Forces that $\rmd V$ and $\rmd V'$ exert on each other point along the line connecting them.

Returning to the main development, \eqref{PDotNewt3} provides an exact expression for the force and torque exerted on an extended mass in Newtonian gravity. It is not, however, particularly useful in this form. It is important to take into account that in many practical scenarios, the effective potential $\hat{\phi}$ (or ``external field'' in this context) varies slowly inside $\Sigma_s$. One might therefore expect that $\hat{\phi}$ could be adequately approximated inside the body using only the first few terms in a Taylor expansion. Integrating each term of such a series recovers standard multipole expansions for the force and torque.

Noting from \eqref{BoxPhiHat} that $\hat{\phi}$ is harmonic in $\Sigma_s$, it must also be analytic in this region (unlike $\phi$, generically). The effective field may therefore be expanded in a Taylor series about, e.g., $\gamma_s \in \Sigma_s$. While it is not guaranteed that the resulting series converges throughout $\Sigma_s$, we assume that it does. The Taylor expansion of $\hat{\phi}$ may be written in a coordinate-invariant manner by introducing Synge's function $\sigma(x,x') = \sigma(x',x)$. This is a two-point scalar equal to one-half of the geodesic distance between its arguments. In Cartesian coordinates $X^i(x)$,
\begin{equation}
	\sigma(x,x') = \frac{1}{2} | \mathbf{X}(x) - \mathbf{X}(x')|^2.
\end{equation}
Derivatives of $\sigma(x, x')$ may be used as ``radial vectors'' between $x$ and $x'$. Holding $x'$ fixed, one derivative of $\sigma(x, x')$ at $x$ produces a 1-form at $x$ whose coordinate components are the ``radial vector''
\begin{equation}
	\sigma_i(x, x') = X_i(x) - X_i(x').
	\label{GradSigma}
\end{equation}
Here, we have used the standard notation $\sigma_a := \nabla_a \sigma$. Using $\LieX \sigma(x,x') = 0$, it follows that
\begin{eqnarray}
	\LieX \hat{\phi}(x') &=& \sum_{n=0}^{\infty} \frac{(-1)^n}{n!} \sigma_{a_1} (\gamma_s, x') \cdots \sigma_{a_n} (\gamma_s, x')
	\nonumber
	\\
	&& ~ \times g^{a_1 b_1}(\gamma_s) \cdots g^{a_n b_n}(\gamma_s) \LieX \nabla_{b_1} \cdots \nabla_{b_n} \hat{\phi} (\gamma_s)
	\label{HatPhiExp}
\end{eqnarray}
for all $x' \in \Sigma_s$ and for all Killing fields $\xi^a$. 

Inserting \eqref{HatPhiExp} into \eqref{PDotNewt3} and integrating term by term,
\begin{equation}
	\frac{\rmd}{\rmd s} P_\xi(s) = - \sum_{n=0}^{\infty} \frac{1}{n!} m^{a_1 \cdots a_n}(s) \LieX \nabla_{a_1} \cdots \nabla_{a_n} \hat{\phi} (\gamma_s),
	\label{PDotNewt4}
\end{equation}
where $m^{a_1 \cdots a_n}$ is the ``complete'' $2^n$-pole mass moment
\begin{eqnarray}
	m^{a_1 \cdots a_n}(s) &:=& (-1)^n g^{a_1 b_1}(\gamma_s) \cdots g^{a_n b_n} (\gamma_s) \nonumber
	\\
	&& ~ \times \int_{\Sigma_s} \rho(x',s) \sigma_{b_1} (\gamma_s, x') \cdots  \sigma_{b_n} (\gamma_s, x') \rmd V'.
	\label{MultipoleDef}
\end{eqnarray}
It is clear that $m^{a_1 \cdots a_n}$ is symmetric in all of its indices. Many of its components do not, however, enter the law of motion \eqref{PDotNewt4}. To see this, note that for any $n \geq 2$, multiples of the (inverse) metric symmetrized with any tensor of rank $n-2$ may be added to $m^{a_1 \cdots a_n}$ without affecting $\rmd P_\xi/\rmd s$. This is a consequence of the fact that $\hat{\phi}$ satisfies the vacuum field equation \eqref{BoxPhiHat}. Using this freedom, each $m^{a_1 \cdots a_n}$ may be replaced by another tensor that is both symmetric and trace-free. For example, the complete quadrupole moment $m^{ab}$ may be replaced by
\begin{equation}
	m^{ab} \rightarrow m^{ab} - \frac{1}{3} g^{ab} g_{cd} m^{cd}.
\end{equation}
The resulting trace-free moments are the ordinary ones found in textbooks. Eq. \eqref{PDotNewt4} is then equivalent to standard multipole expansions for the force and torque acting on an extended mass in Newtonian gravity. 

More explicit equations may be obtained by fixing the point $\gamma_s$. It is natural to do so by choosing this to lie at the body's center of mass. The center of mass is defined to be the point about which the first mass moment vanishes:
\begin{equation}
	m^a(s) = - g^{ab}(\gamma_s)  \int_{\Sigma_s} \rho(x',s) \sigma_b (\gamma_s, x') \rmd V' = 0.
	\label{CMNewt}
\end{equation}
A unique solution to this equation is guaranteed by the assumption that the mass density can never be negative and $m \neq 0$. Choosing $\gamma_s$ such that $m^a=0$ eliminates the dipole ($n=1$) term in \eqref{PDotNewt4}. Differentiating \eqref{CMNewt} with respect to $s$ and using \eqref{MassCons} also demonstrates that
\begin{equation}
	p^a = m \dot{\gamma}^a_s.
	\label{MomVelNewt}
\end{equation}
This is the ordinary relation between linear momentum and center of mass velocity. Note, however, that a similar equation does not remain true in the relativistic case (although it is often an excellent approximation). 

Explicit laws of motion may now be written down for the center of mass position $\gamma_s$, linear momentum $p^a$, and spin $S_a$. Combining \eqref{StoSTensNewt}, \eqref{PDotFNNewt}, \eqref{PDotNewt4}, and \eqref{CMNewt} shows that
\begin{eqnarray}
	\fl \qquad \frac{ \rmd p^a }{ \rmd s } = - g^{ab}(\gamma_s) \left( m \nabla_b \hat{\phi}(\gamma_s) + \sum_{n=2}^\infty \frac{1}{n!} m^{c_1 \cdots c_n}(s) \nabla_{b} \nabla_{c_1}  \cdots \nabla_{c_n} \hat{\phi} (\gamma_s) \right) ,
\label{NewtMultF}
\\
	\fl \qquad \frac{ \rmd S_a }{ \rmd s} = - \epsilon_{a b_1 c} g^{cd}(\gamma_s) \sum_{n=2}^\infty \frac{1}{(n-1)!} m^{b_1 \cdots b_n}(s) \nabla_d \nabla_{b_2} \cdots \nabla_{b_n} \hat{\phi} (\gamma_s) . 
\label{NewtMultN}
\end{eqnarray}
Combining the first of these equations with \eqref{MomVelNewt} and $\rmd m /\rmd s =0$ immediately provides a similar expansion for the acceleration of a body's center of mass in terms of its multipole moments and derivatives of $\hat{\phi}$ evaluated at $\gamma_s$. 

In most cases of practical interest, the first few terms in \eqref{NewtMultF} and \eqref{NewtMultN} provide excellent approximations to the true force and torque. If, e.g., the object of interest has a size $O(d)$ and is separated from other objects by a distance of $O(D)$, successive terms in the multipole expansions tend to differ in size by a factor of at least $d/D$. A more precise bound may be obtained using standard expressions for the remainder term associated with a Taylor series of finite order. Even better estimates can be found using Fourier transforms. See, e.g., \cite{Dix67}.

Note that the gravitational potential $\hat{\phi}$ entering into the final laws of motion is not the one that would be measured using local experiments (which is $\phi$, or really its gradient). The body's momenta satisfy evolution equations that are instantaneously identical to those of an extended test mass with moments $m^{a_1 \cdots a_n}$ moving in the vacuum field $\hat{\phi}$ ($\neq \phi$).

It is the intent of this paper to demonstrate a similar result for self-gravitating masses in general relativity. This is done in two steps. First, Sect. \ref{Sect:TestGR} considers relativistic test masses moving in a prescribed spacetime. The laws of local mass and momentum conservation \eqref{MassCons} and \eqref{MomentumCons} are then replaced by conservation of the body's stress-energy tensor. No particular relation is, however, assumed to hold between the body of interest and the spacetime metric. Once the relevant techniques are established, Einstein's equation is ``turned on'' in the remainder of this paper and self-interaction is dealt with directly.


\subsection{Test masses in curved spacetimes}
\label{Sect:TestGR}

Consider a relativistic extended body moving in a curved four-dimensional spacetime $(\mathcal{M}, g_{ab})$. Associated with this body is a stress-energy tensor $T^{ab} = T^{(ab)} (x)$. Denoting its worldtube by $W := \mathrm {supp} \, \, T^{ab}$, spatial slices of $W$ are assumed to be compact and to have positive (but finite) 3-volume. As in the Newtonian case discussed above, all laws of motion are to be derived from generic local conservation laws. In this context, Eqs. \eqref{MassCons} and \eqref{MomentumCons} are replaced by stress-energy conservation:
\begin{equation}
	\nabla_{a} T^{ab} =0.
\label{StressCons}
\end{equation}	
We assume for now that the body of interest is a test mass, meaning that it does not affect the spacetime metric $g_{ab}$. There is therefore no replacement for \eqref{Poisson} in this section.

Now consider the ``momentum functional''
\begin{equation}
	\ItP_\xi(\Sigma) := \int_\Sigma g_{ab}  \xi^{a} T^{bc} \rmd S_c,
	\label{BarePDef}
\end{equation}
where $\rmd S_c$ denotes the natural 3-surface element associated with $g_{ab}$. $\ItP_\xi(\Sigma)$ takes as input a hypersurface $\Sigma$ assumed to bisect $W$ and a vector field $\xi^{a}$ that is chosen later. As with the similar functional \eqref{PXiNewt} defined in the Newtonian case, $P_\xi(\Sigma)$ may be viewed as returning the component of momentum conjugate to $\xi^{a}$ at a ``time'' defined by $\Sigma$. This interpretation is completely standard if $\zeta^{a}$ is a Killing vector: Eq. \eqref{StressCons} then implies that $\ItP_{\zeta} (\Sigma)$ is independent of $\Sigma$ (i.e., it is conserved). 

As in the Newtonian case, linear and angular momenta $p^a$ and $S^{ab} = S^{[ab]}$ may be defined by demanding that $P_\xi$ be a linear combination of these two quantities. Specifically, it is useful to retain \eqref{PtopSNewt}. This relation does not, however, make sense without being more specific about the types of vector fields $\xi^a$ that may be used in \eqref{BarePDef}. In general, there is no reason to expect that any Killing fields exist (and certainly not the $4+6 = 10$ required to define all components of $p^a$ and $S^{ab}$). Using a relation like \eqref{PtopSNewt} requires that $\xi^a$ be chosen from a ten-dimensional vector space with the property that each vector is determined throughout a hypersurface $\Sigma_s$ given knowledge of an arbitrary 1-form $\Xi_a(\gamma_s)$ and an arbitrary 2-form $\nabla_a \Xi_b = \nabla_{[a} \Xi_{b]} (\gamma_s)$ at one point $\gamma_s \in \Sigma_s$. Furthermore, $\xi^a(x)$ must be linear in this ``initial data.'' 

There are many spaces of vector fields with these properties. We now specialize to specific definitions that recover Dixon's definitions \cite{Dix79, Dix74, Dix70a} for the linear and angular momentum of an extended body. Using the terminology of \cite{HarteSyms}, $\xi^{a}$ is assumed to be of the form\footnote{The simpler notation $\xi^{a} = g^{ab} \xi_{b}$ is not used in order to avoid confusion when multiple metrics are introduced below.}
\begin{equation}
	\xi^{a} = g^{ab} \Xi_{b},
	\label{OmegaToXi}
\end{equation}
where $\Xi_{a}$ is a Killing-type generalized affine collineation constructed using $g_{ab}
$. This is defined precisely in the appendix. Following \cite{HarteEMNew, HarteHighMult, HarteScalar}, we simplify the terminology by referring to the $\Xi_{a}$ (or $\xi^{a}$) simply as generalized Killing fields (GKFs) with respect to $g_{ab}$. 

\begin{figure}
	\centering
	\includegraphics[width= .30 \linewidth]{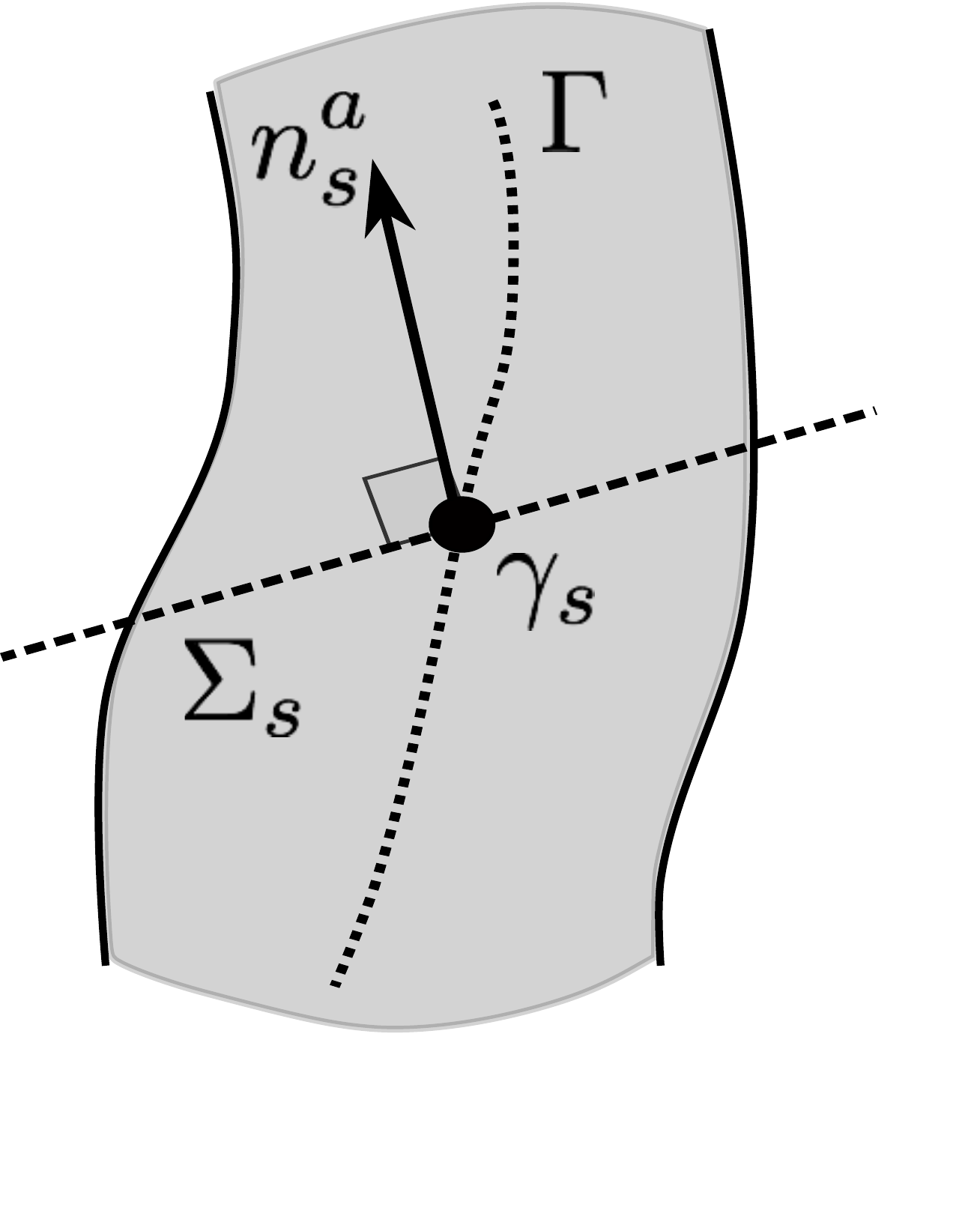}
	\vspace{-.8 cm}
	\caption{A schematic illustration of the geometry required to define generalized Killing fields: The shaded region denotes a portion of the body's worldtube $W$. $\gamma_s$ is a point on the timelike worldline $\Gamma$. $n^a_s$ is a timelike vector at $\gamma_s$ and $\Sigma_s$ is a spacelike hypersurface formed by the union of all geodesics orthogonal to $n^a_s$.}
	\label{Fig:geometry}
\end{figure}

Defining GKFs requires fixing not only a metric, but also a timelike worldline $\Gamma = \{ \gamma_{s} | s \in \mathbb{R} \}$ and a timelike vector field $n^{a}_{s} \in T_{\gamma_{s}} \mathcal{M}$ along $\Gamma$. The worldline serves as an origin about which to compute multipole moments of $T^{ab}$. The $n^{a}_{s}$ fix a family of spacelike hypersurfaces $\Sigma_{s}$ that provide a time function $\Sigma_{s} \ni x \mapsto s$ inside the body's worldtube $W$. At fixed $s$, $\Sigma_{s}$ is defined to be the union of all geodesics that pass through $\gamma_{s}$ and are orthogonal to $n^{a}_{s}$ at that point. These geodesics are not to be extended so far that they intersect either with each other (except at $\gamma_{s}$) or with another hypersurface in the family. It is assumed that the body is sufficiently small that such restricted geodesics still form hypersurfaces $\Sigma_{s}$ that foliate $W$. See Fig. \ref{Fig:geometry} for an illustration of the geometry. Under mild assumptions, $\gamma_{s}$ and $n^{a}_{s}$ can both be specified uniquely using center of mass conditions \cite{CM} (see also \eqref{pPropN} and \eqref{pDotS} below). For now, however, we continue to describe the general case where they are left free.

Once $g_{ab}$, $\Gamma$, and $n^{a}_{s}$ have been fixed, vector fields $\xi^{a} = g^{ab} \Xi_b$ that may be used in $P_\xi$ are to be chosen using the definitions in the appendix. The result is a ten-dimensional vector space with a number of characteristics that are very similar to those of genuine Killing fields. First among these is the ``rigidity property'' that has already been mentioned: Given any 1-form $\Xi_{a}(\gamma_{s})$ and 2-form $\nabla_{a} \Xi_{b} = \nabla_{[a} \Xi_{b]}(\gamma_{s})$ at a single point $\gamma_{s} \in \Gamma$, a GKF $\Xi_{a}(x)$ is fixed for all $x$ in the neighborhood $\mathcal{W}$ of $\Gamma$ defined in the appendix. The $\Xi_a(x)$ are linear in $\Xi_a(\gamma_s)$ and $\nabla_a \Xi_b (\gamma_s)$.

GKFs are also ``approximately Killing'' near $\Gamma$, meaning that
\begin{equation}
	\nabla_{(a} \Xi_{b)}|_{\Gamma} = \nabla_{a} \nabla_{(b} \Xi_{c)}|_{\Gamma} = 0,
\end{equation}
or equivalently,
\begin{equation}
	\Lie_{\xi} g_{ab}|_{\Gamma} = \nabla_{a} \Lie_{\xi} g_{bc}|_{\Gamma} = 0.
	\label{NearlyKilling}
\end{equation}
The space of generalized Killing fields includes any genuine Killing fields associated with $g_{ab}$: If $\zeta^{a}$ satisfies $\Lie_{\zeta} g_{ab} =0$ everywhere, it is also a generalized Killing field.  In maximally-symmetric spacetimes, the space of generalized Killing fields coincides with the space of genuine Killing fields. The dependence on a preferred worldline and foliation disappears in this special case. More generally, the $\xi^a$ may be interpreted as the generators of an ``approximate Poincar\'{e} group'' for an observer moving on $\Gamma$ (and with a preferred time-slicing determined by the $\Sigma_s$).

Linear and angular momenta $p^{a}(s)$ and $S^{ab} = S^{[ab]}(s)$ may now be introduced as tensor fields along $\Gamma$. Following \eqref{PtopSNewt}, let
\begin{equation}
	\ItP_{\xi}(\Sigma_{s}) = p^{a} (s) \Xi_{a} (\gamma_{s}) + \frac{1}{2} S^
	{ab} (s) \nabla_{a} \Xi_{b} (\gamma_{s}).
	\label{pSFormula} 
\end{equation}
As in the Newtonian case, knowledge of $\ItP_{\xi}$ for all possible $\xi^{a}$ is equivalent to knowledge of $p^{a}$ and $S^{ab}$. Here, however, $S^{ab}$ is not equivalent to an angular momentum 1-form. This second-rank tensor has six independent components in the four spacetime dimensions considered here. Roughly speaking, there is a sense in which $S^{ab}$ contains information associated with both the spin 1-form and the mass dipole moment of Newtonian physics.

It is possible to write $\Xi_a(x)$ directly in terms of $\Xi_a(\gamma_s)$ and $\nabla_a \Xi_b(\gamma_s)$ if $x \in \Sigma_s$. The resulting expressions use Synge's function $\sigma(x,x')$. Recall that this is a biscalar defined to equal one-half of the geodesic distance between its arguments. To borrow terminology from optics, it is essentially a characteristic function for the spacetime. Many properties of $\sigma(x,x')$ are discussed in \cite{PoissonRev, Synge, Friedlander}. As is standard, we denote derivatives of $\sigma$ by appending indices: e.g., $\sigma_{aa'} := \nabla_a \nabla_{a'} \sigma$. Defining $H^{a'a}(x,x') := [-\sigma_{a a'} (x,x') ]^{-1}$, and $K^{a'}{}_{a}(x,x') :=  H^{a'b}(x,x') \sigma_{ab} (x,x')$, all GKFs may be shown to satisfy \cite{HarteSyms}
\begin{equation}
	\fl \qquad \xi^{a'} (x') = g^{ab}(\gamma_s)  \big[ K^{a'}{}_{a} (\gamma_s,x') \Xi_{b}(\gamma_s) - H^{a'c} (\gamma_s, x') \sigma_a (\gamma_s, x') \nabla_b \Xi_c (\gamma_s) \big]
\end{equation}
if $x'$ lies within the normal neighborhood of $\gamma_s$ and $x' \in \Sigma_s$. This equation generalizes the Euclidean expression \eqref{EuclKilling}. Inserting it into \eqref{BarePDef} and \eqref{pSFormula} provides explicit expressions for the momenta as integrals over the body's stress-energy tensor:
\begin{equation}
	p^{a}(s) = g^{ab}(\gamma_s) \int_{\Sigma_{s}} g_{a'b'}(x') K^{a'}{}_{b}(\gamma_{s},x') T^{b'c'}(x') \rmd S_{c'},
	\label{pExpl}
\end{equation}
and
\begin{equation}
	\fl \quad \qquad S^{ab}(s) = 2 \int_{\Sigma_{s}} g_{a'b'}(x')  H^{a'[a} (\gamma_{s},x') g^{b]c} (\gamma_{s}) \sigma_c (\gamma_{s},x') T^{b'c'}(x') \rmd S_{c'}.
	\label{SExpl}
\end{equation}
These momenta coincide with standard textbook definitions in flat spacetime. In curved spacetimes, they are the momenta identified by Dixon as being particularly useful for the description of objects with conserved stress-energy tensors \cite{Dix79,Dix74,Dix70a}. Formulae \eqref{pExpl} and \eqref{SExpl} are included here for completeness, but are not needed in any arguments below.

As in Newtonian physics, changes in the momenta may be computed from changes in $P_\xi$. First note that \eqref{NearlyKilling} may be used to show that for any GKF $\Xi_a$ and any $\gamma_s \in \Gamma$ \cite{HarteSyms},
\begin{equation}
	\nabla_b \nabla_a \Xi_c(\gamma_s) = R_{cab}{}^{d}(\gamma_s) \Xi_d(\gamma_s).
	\label{Del2Xi}
\end{equation}
An equivalent relation holds everywhere for genuine Killing fields \cite{Wald}. Differentiating \eqref{pSFormula} while using this identity,
\begin{equation}
	\fl \qquad \qquad \frac{ \rmd }{\rmd s} \ItP_{\xi}  = \left( \frac{ \rmD p^{a} }{\rmd s} - \frac{1}{2} R_{bcd}
	{}^{a} S^{bc} \dot{\gamma}_{s}^{d} \right) \Xi_{a} + \frac{1}{2} \left( \frac{ \rmD S^		{ab} }{\rmd s} - 2 p^{[a} \dot{\gamma}^{b]}_{s} \right) \nabla_{a} \Xi_{b}. 
	\label{PDotGenTest}
\end{equation}
As is standard, the notation $\dot{\gamma}^{a}_{s}$ used here denotes the tangent vector to the curve $\Gamma$ at $\gamma_s$. Eq. \eqref{PDotGenTest} provides a recipe for extracting the covariant derivatives $\rmD p^{a}/\rmd s$ and $\rmD S^{ab}/\rmd s$ from $\rmd \ItP_{\xi}(\Sigma_{s})/\rmd s$. The only difference between this equation and its Newtonian equivalent \eqref{PDotFNNewt} is the presence of the Riemann tensor $R_{bcd}{}^{a}$. This arises from \eqref{Del2Xi}.

Everything said thus far has involved only definitions. It is now possible to explore the physical properties of the momenta that have just been described. In general, $\ItP_{\xi} (\Sigma_s)$ depends on $s$. Using \eqref{StressCons} and \eqref{BarePDef}, the difference in $P_\xi$ between two times $s$ and $s'>s$ is
\begin{eqnarray}
	\delta \ItP_\xi(\Sigma_{s},\Sigma_{s'}) &:=& \ItP_\xi(\Sigma_{s'}) - \ItP_\xi(\Sigma_{s})
	\nonumber
	\\
	&=& \frac{1}{2} \int_{\Omega(s,s')}  \! \! T^{ab} \LieX g_{ab}  \rmd V,
\label{DeltaPTest}
\end{eqnarray}
where $\Omega(s,s')$ is defined to be the portion of the body in between the two hypersurfaces $\Sigma_{s}$ and $\Sigma_{s'}$. $\rmd V$ denotes the natural (four-dimensional) volume element associated with $g_{ab}$. Eq. \eqref{DeltaPTest} may be put into differential form by letting $\rmd S := t^{a} \rmd S_{a}$, where $t^{a}$ is a time evolution vector field for the foliation $\{ \Sigma_{s} \}$:
\begin{equation}
	\frac{\rmd}{\rmd s} \ItP_{\xi} (\Sigma_{s}) =  \frac{1}{2} \int_{\Sigma_{s}} 	T^{ab} \LieX g_{ab} \rmd S.
\label{DeltaPTestInst}
\end{equation}
Equating the right-hand side of this equation with the right-hand side of \eqref{PDotGenTest} provides evolution equations for $p^{a}$ and $S^{ab}$.

It is clear from \eqref{DeltaPTest} that $P_\xi$ is a conserved quantity if $\xi^a$ is Killing. If $\rmd \ItP_{\xi} / \rmd s = 0$ for all $\xi^a$, one recovers the Papapetrou equations \cite{Mathisson, Papapetrou} typically used to model a spinning test particle. More generally, changes in $\ItP_{\xi}$ measure the deviation from these equations. In this formalism, Papapetrou terms in the laws of motion arise purely as a kinematic consequence of \eqref{NearlyKilling} and \eqref{pSFormula}.  

The discussion up to this point has not made any strong assumptions regarding the nature of the metric. In particular, self-fields have not been excluded. We now assume, however, that in a Riemann normal coordinate system $X^\mu(x)$ with origin $\gamma_{s}$, the metric components $g_{\mu\nu}$ may be accurately expanded throughout $\Sigma_{s} \cap W$ in a Taylor series about $\gamma_{s}$. In particular, introduce four 1-forms $e^\mu_a$ at $\gamma_s$. These are assumed to form an orthonormal tetrad, so
\begin{equation}
	g^{ab} (\gamma_s) e^\mu_a e^\nu_b  = \eta^{\mu\nu} , \qquad \eta_{\mu\nu}  e^\mu_a e^\nu_b  = g_{ab}(\gamma_s), 
\end{equation}
where $\eta_{\mu\nu} = \eta^{\mu\nu} = \mathrm{diag}(-1,1,1,1)$. This tetrad allows the introduction of four Riemann normal coordinates $X^\mu (x')$ associated with the point $x'$:
\begin{equation}
	X^\mu (x') := - e_a^\mu g^{ab}(\gamma_s) \sigma_b(\gamma_s, x').
\end{equation}
This definition is, in part, motivated by the Euclidean expression \eqref{GradSigma}. Note that $X^\mu(\gamma_s) = 0$, so $\gamma_s$ is the origin of this coordinate system.

The metric in Riemann normal coordinates can be viewed as a matrix of scalars that depend on the choice of origin $\gamma_s$ and the coordinates $X^\mu(x')$. This matrix is given by \cite{HarteHighMult}
\begin{equation}
	g_{\mu\nu} (\gamma_s,X^\mu(x') ) = \eta_{\mu\lambda} \eta_{\nu\rho} e^\lambda_a e_b^\rho H^{a'a} (\gamma_s,x') H^{b'b} (\gamma_s,x') g_{a'b'} (x') .
\end{equation}
Taylor expanding these scalars\footnote{In general, $g_{\mu\nu}$ need not be analytic at $\gamma_s$. We assume that a finite power series nevertheless provides an adequate approximation throughout $\Sigma_s \cap W$. The Taylor expansion \eqref{MetricExpand} should therefore be cut off at finite $n$. We write an infinite upper limit and an exact equality sign here (and in similar equations below) for simplicity.}  in $X^\mu(x')$ about $X^\mu = 0$ leads to a general expression for the metric that does not make any explicit reference to the $e^\mu_a$. Letting $X^a (\gamma_s,x'):= - g^{ab}(\gamma_s) \sigma_b(\gamma_s,x')$, the resulting series is \cite{HarteHighMult}
\begin{eqnarray}	
	g_{a'b'}(x') = \sigma_{aa'}(\gamma_s, x') \sigma_{bb'}(\gamma_s, x') g^{ac}(\gamma_s) g^{bd}(\gamma_s)  
	\nonumber
	\\
 	\qquad ~ \times \sum_{n=0}^{\infty} \frac{1}{n!} X^{f_1 }(\gamma_s,x') \cdots X^{f_n} (\gamma_s, x') g_{cd,f_1 \cdots f_n}(\gamma_s).
 	\label{MetricExpand}
\end{eqnarray}
The $g_{ab,c_1 \cdots c_n}(\gamma_s)$ appearing in this equation are referred to either as tensor extensions of $g_{ab}$ or as metric normal tensors. They are derived from the coefficients appearing in the Taylor series for $g_{\mu\nu}$:
\begin{equation}
	g_{ab,c_1 \cdots c_n} (\gamma_s) := e_a^\mu e_b^\nu e_{c_1}^{\lambda_1} \cdots e_{c_n}^{\lambda_n}  \frac{ \partial^n g_{\mu\nu}(\gamma_s,0) }{ \partial X^{\lambda_1} \cdots \partial X^{\lambda_n} } .
\end{equation}
Despite appearances, the $g_{ab,c_1 \cdots c_n}$ do not depend on the choice of $e_a^\mu$. The zeroth extension is the metric itself and the first extension vanishes. In general, it is clear that the $n^{\mathrm{th}}$ metric normal tensor is symmetric in both its first two and its last $n$ indices. It may also be shown that \cite{HarteHighMult}
\begin{equation}
	g_{a(b,c_{1} \cdots c_{n})} = g_{(ab,c_{1} \cdots c_{n-1}) c_{n}} = 0
\end{equation}
for all $n \geq 2$. Keeping this restriction on $n$, all metric normal tensors can be written as polynomials in the Riemann tensor. To linear order \cite{Dix74},
\begin{equation}	
	g_{ab,c_{1} \cdots c_{n}} = 2 \left( \frac{n-1}{n+1} \right) \nabla_{(c_{3} \cdots c_{n}} (R_{|a|c_{1} c_{2})}{}^{d} g_{bd}) + O(R^{2}).
\label{MetricExtensionApprox}
\end{equation}
This equation is exact for $n=2,3$. For higher $n$, there are additional terms nonlinear in $R_{abc}{}^{d}$ or its derivatives. 

Using certain details of the GKFs together with \eqref{MetricExpand}, one may derive a power series expansion for $\LieX g_{ab}$ \cite{HarteHighMult}:
\begin{eqnarray}
	\LieX g_{a'b'}(x') &=& \sum_{n=2}^{\infty} \frac{1}{n!} (\cdots)_{a'b' d_1 \cdots d_n}{}^{ab c_1 \cdots c_n} 
	\nonumber
	\\
	&& ~ \times X^{d_1} (\gamma_s, x') \cdots X^{d_n} (\gamma_s, x') \LieX g_{ab,c_1 \cdots c_n} (\gamma_s).
	\label{LieGExpand}
\end{eqnarray}
The omitted coefficients in this series are known explicitly in terms of $\sigma$ if $x' \in \Sigma_s$, which is the only case relevant in this section. More generally, such a series still exists, although the coefficients are no longer known exactly. Substituting \eqref{LieGExpand} into \eqref{DeltaPTestInst} now shows that
\begin{equation}
	\frac{\rmd}{\rmd s} \ItP_{\xi}(\Sigma_{s}) = \frac{1}{2} \sum_{n=2}^{\infty} \frac{1}{n!} I^{c_{1} \cdots c_{n} a b}(s) 	\LieX g_{ab, c_{1} \cdots c_{n}} (\gamma_{s}).
	\label{TestBodyForceExp}
\end{equation} 
By analogy with \eqref{PDotNewt4}, the coefficients $I^{c_{1} \cdots c_{n} ab}(s)$ appearing here are interpreted as the $2^{n}$-pole moments of $T^{ab}$ at the time $s$. 

Without loss of generality, the symmetry properties of the metric normal tensors allow the $I^{c_{1} \cdots c_{n} ab}$ to be chosen such that they are separately symmetric in their first $n$ and last two indices. They may also be taken to satisfy
\begin{eqnarray}
	I^{(c_{1} \cdots c_{n} a)b } = I^{c_{1} (c_{2} \cdots c_{n} ab)} = 0.
	\label{ISym}
\end{eqnarray}
A unique formula linking moments with these properties to $T^{ab}$ may be derived using \eqref{LieGExpand} and \eqref{TestBodyForceExp} \cite{HarteHighMult} (see also \cite{Dix74}). Like \eqref{pExpl} and \eqref{SExpl}, the result has the form of an integral over $\Sigma_{s}$ involving the stress-energy tensor and various bitensors constructed from $\sigma$. It is significantly more complicated than the Newtonian formula \eqref{MultipoleDef} for $m^{a_1 \cdots a_n}$. This is partially because $T^{ab}$ has two more indices than $\rho$. Much less obvious is that the relativistic moments are ``reduced'' with respect to \eqref{StressCons}. They are adapted to describing \textit{conserved} second-rank symmetric tensors. Knowing all of the $I^{c_1 \cdots c_n ab}$ together with $p^{a}$ and $S^{ab}$ is equivalent to knowledge of $T^{ab}$  \cite{Dix74}. The same statement does not remain true if $T^{ab}$ is replaced in all integrals by a second-rank symmetric tensor that is not divergence-free.

The given index symmetries imply that $I^{c_{1} \cdots c_{n} ab}$ has a total of
\begin{equation}
	(n+3)(n+2)(n-1)
	\label{MomentComps}
\end{equation}
algebraically independent components. This far exceeds the number typically ascribed to the $2^n$-pole moment in other formalisms \cite{ThorneMoments}. The reason for this is essentially that the $I^{\cdots}$ are ``complete'' in the sense described in the previous paragraph. If no restrictions are placed on $g_{ab}$, nothing further can be said. Recall, however, that traces of $m^{a_1 \cdots a_n}$ decouple from the Newtonian equation \eqref{PDotNewt4} because $\hat{\phi}$ is a vacuum field. Similarly, certain components of $I^{\cdots}$ decouple from \eqref{TestBodyForceExp} if $g_{ab}$ satisfies the vacuum Einstein equation $R_{ab} = 0$. This may be seen by noting that certain traces of \eqref{MetricExtensionApprox} vanish in this case. Use of \eqref{NearlyKilling} shows that these same traces still vanish if $R_{ab} = \Lambda g_{ab}$ for any constant $\Lambda$. In most cases where a test body description is appropriate, the $I^{\cdots}$ may therefore be replaced in \eqref{TestBodyForceExp} by moments with many fewer components. Additional discussion of these points may be found in \cite{HarteHighMult}, although precise details of the reduction process are not known. 

Another important point to note is that the sum in \eqref{TestBodyForceExp} starts at $n=2$. This corresponds to quadrupole order. It is a consequence of \eqref{NearlyKilling} and \eqref{DeltaPTestInst} that the monopole and dipole moments of $T^{ab}$ -- essentially $p^{a}$ and $S^{ab}$ -- do not directly contribute to $\rmd \ItP_{\xi} / \rmd s$. These moments do, however, affect $\rmD p^a/\rmd s$ and $\rmD S^{ab}/\rmd s$ via the Papapetrou-like terms appearing in \eqref{PDotGenTest}. Explicitly, define a net force $F^{a}(s)$ and a net torque $N^{ab} = N^{[ab]}(s)$ such that
\begin{eqnarray}
	\frac{\rmD p^{a}}{\rmd s} = \frac{1}{2} R_{bcd}{}^{a} S^{bc} \dot{\gamma}^{d}_{s} + F^{a},
	\label{ForceDefTest}
	\\
	\frac{ \rmD S^{ab} }{\rmd s} = 2 p^{[a} \dot{\gamma}^{b]}_{s} + N^{ab}.
	\label{TorqueDefTest}
\end{eqnarray} 
Comparison with \eqref{pSFormula} and \eqref{TestBodyForceExp} shows that
\begin{equation}
	F^{a}(s) = \frac{1}{2} g^{ab}(\gamma_{s}) \sum_{n=2}^{\infty} \frac{1}{n!} I^{f_{1} \cdots 	f_{n} cd}(s) \nabla_{b} g_{cd, f_{1} \cdots f_{n}}(\gamma_{s}),
	\label{ForceMultTest}
\end{equation}
and
\begin{eqnarray}
	\fl N^{ab}(s) =2 \sum_{n=2}^{\infty} \frac{1}{n!} I^{c_{1} \cdots c_{n} df}(s) \Big[ g_{fh, 	c_{1} \cdots c_{n} } (\gamma_{s}) \delta^{[a}_{d} + \frac{n}{2} g_{df, h c_{1} \cdots c_{n-1}} (\gamma_{s}) \delta^{[a}_{c_{n}} \Big] g^{b]h} (\gamma_{s}).
	\label{TorqueMultTest}
\end{eqnarray}

The hope in writing these series is, of course, that adequate approximations may be obtained by truncating them at some small maximum $n$. This can only happen if $\Gamma$ and $\{\Sigma_{s}\}$ are chosen appropriately (if it is possible at all for a given system). We now fix a particular worldline and foliation that is hopefully ``appropriate'' in this sense. This is done by imposing center of mass conditions as described in, e.g., \cite{Dix79, Dix70a,EhlRud}. First recall that $\Sigma_{s}$ is constructed using geodesics that pass through $\gamma_{s}$ and are orthogonal to $n^{a}_{s}$ at that point. Suppose that $\Gamma$ and $n^{a}_{s}$ are chosen such that
\begin{eqnarray}
	p^{a}(s) \propto n^{a}_{s},
	\label{pPropN}
	\\
	g_{ab}(\gamma_{s}) p^{a}(s) S^{bc}(s) =0.
	\label{pDotS}
\end{eqnarray}
Under mild assumptions, the resulting $\Gamma$ and $n^{a}_{s}$ exist, are unique, and are timelike \cite{CM}. The first of these equations essentially states that the $\{\Sigma_s\}$ foliation is the one preferred by ``zero-momentum observers.'' Eq. \eqref{pDotS} encapsulates the notion that a body's center of mass position is the point about which its mass dipole moment vanishes in the zero-momentum frame.

Unlike in Newtonian gravity, the center of mass velocity $\dot{\gamma}^{a}_{s}$ of a relativistic mass is not necessarily proportional to $p^{a}$. Relating these two quantities is simpler if the time parameter $s$ is chosen such that $g_{ab}(\gamma_{s}) p^{a}(s) \dot{\gamma}^{b}_{s} = - m(s)$, where the mass is defined by
\begin{equation}
	m(s) := [ - g_{ab}(\gamma_s) p^{a}(s) p^{b}(s)]^{1/2} .
\end{equation}
This means that in general, $\dot{\gamma}^{a}_{s}$ does not have unit norm. There is, however, no loss of generality in assuming that $g_{ab} n^{a}_{s} n^{b}_{s} = -1$. Hence,
\begin{equation}
	p^{a} = m n^{a}_{s}.
\end{equation}
With these conventions, differentiation of \eqref{pDotS} may be used to show that the linear momentum and center of mass velocity are related via \cite{EhlRud}
\begin{equation}
	\fl \qquad m \dot{\gamma}^{a}_{s} = p^{a} - N^{ab} g_{bc} n^{c}_{s} - \frac{ S^{ab} [m g_{bc} F^{c} - \frac{1}{2} S^{cd} (p^{f} - N^{fh} g_{hr} n^{r}_{s})  R_{cdb}{}^{l} g_{fl} ] }{m^{2} + \frac{1}{4} S^{bc} S^{df} R_{bcd}{}^{l} g_{fl}} .
	\label{CMVelTest}
\end{equation}
This equation breaks down if $m^{2} + \frac{1}{4} S^{bc} S^{df} R_{bcd}{}^{l} g_{fl} = 0$, which may be interpreted as a constraint on $S^{ab}/m$. Such a restriction is implied by the conditions required for the center of mass to exist as a unique timelike worldline.

Note that the center of mass velocity does not appear on the right-hand side of \eqref{CMVelTest}. Indeed, the complexity of this equation arises mainly from the nontrivial operations required to solve explicitly for $\dot{\gamma}^a_s$. Displaying \eqref{CMVelTest} in this way makes it evident that Eqs.  \eqref{ForceDefTest}-\eqref{TorqueMultTest} and \eqref{CMVelTest} form ordinary differential equations (ODEs) for $\gamma_{s}$, $p^{a}$, and $S^{ab}$ with the form
\begin{equation}
	\dot{\gamma}_s^a = \ldots , \qquad \frac{ \rmD p^a }{\rmd s} = \ldots , \qquad \frac{ \rmD S^{ab} }{\rmd s} = \ldots .
\end{equation}
The right-hand sides of these equations involve only geometric quantities, $\gamma_s$, $p^a$, $S^{ab}$, and the higher moments $I^{\cdots}$. It follows that if the metric is known and the quadrupole and higher moments are prescribed functions of $s$, the motion is uniquely determined by specifying initial values for $\gamma_s$, $p^a$, and $S^{ab}$. There is, however, no physical reason that the higher moments should be treated as given functions of $s$. Generically, their evolution depends on the details of the specific system under consideration. 

To summarize, Dixon's momenta have now been described as they apply to a compact body with a conserved stress-energy tensor. Eq. \eqref{BarePDef} constructs $P_\xi$ from $\xi^a$ and $T^{ab}$, \eqref{pSFormula} links $p^a$ and $S^{ab}$ to $P_\xi$, and the $\xi^a$ are built from the metric in the appendix. Taken together, these relations define Dixon's momenta in a very general context. Their changes are described exactly by the integral \eqref{DeltaPTestInst} (even if the body's presence strongly influences the geometry). As in Newtonian gravity, integral expressions for the force and torque are not particularly useful by themselves. Ideally, one would like to be able to constrain the motion without detailed knowledge of the body's stress-energy tensor. This is easily accomplished in the formalism just described if the metric does not vary rapidly inside the body (specifically, on a given $\Sigma_{s} \cap W$). $\rmd \ItP_{\xi} / \rmd s$ can then be expanded in the multipole series \eqref{TestBodyForceExp}. This is, however, a reasonable procedure mainly for test masses. In cases involving significant self-gravity, $g_{ab}$ generically varies rapidly inside the body. In such cases, it cannot be accurately approximated using a low-order Taylor series.

Nevertheless, the remainder of this paper establishes that there \textit{is} a sense in which useful multipole expansions can be performed even in the presence of a significant self-field. This is because forces and torques exerted by the (potentially complicated) self-field are shown only to make an object's momenta and higher moments appear slightly shifted from what might have initially been expected. In this sense, the self-field may be eliminated from the instantaneous laws of motion by appropriate (and physically reasonable) redefinitions.

Obtaining this result requires defining precisely what ``should'' be meant by ``linear and angular momentum'' as well as ``self-field.'' The momenta used below reduce to the definitions provided in this section in a test mass limit. More than this, the form \eqref{BarePDef} for $\ItP_{\xi}(\Sigma)$ is retained exactly as written. The space of vector fields from which the $\xi^a$ are to be drawn does change, however.

\section{Mechanics of self-gravitating extended bodies in curved spacetimes}
\label{Sect:MechanicsGen}

The remainder of this paper considers an extended mass with stress-energy tensor $T^{ab}$ moving in a spacetime $(\mathcal{M},g_{ab})$. Its worldtube is denoted by $W$. A worldline $\Gamma$ is taken as given together with a collection of hypersurfaces $\{ \Sigma_s | s \in \mathbb{R} \}$ that foliate $W$. Each slice $\Sigma_s \cap W$ is assumed to be compact and $\gamma_s := \Gamma \cap \Sigma_s$ (see Fig. \ref{Fig:geometry}). Both $\Gamma$ and $\{ \Sigma_s\}$ may be fixed using center of mass conditions (see Sect. \ref{Sect:CM} below), although we do not require this. Unlike in Sect. \ref{Sect:TestGR}, Einstein's equation 
\begin{equation}
	R_{ab} - \frac{1}{2} g_{ab} R + \Lambda g_{ab}  = 8 \pi g_{ac} g_{bd}T^{cd}  
\label{Einstein}
\end{equation}
is now required to hold with stress-energy tensor $T^{ab}$ at least in a neighborhood of $W$ (i.e., the body is assumed not to be in direct contact with any other source of stress-energy). The presence of a cosmological constant $\Lambda$ in Einstein's equation does not significantly change any arguments below, so we allow it to be nonzero. Additionally, note that $T^{ab}$ is automatically conserved as an integrability condition for \eqref{Einstein}.

The ``bare'' momentum associated with a body of the type just described is defined by introducing a $P_\xi (\Sigma_s)$ computed via \eqref{BarePDef} as a certain integral of $T^{ab}$ over $\Sigma_s$. All metrics and volume elements in that formula are those associated with the physical metric $g_{ab}$. This is to be distinguished from a certain ``effective metric'' $\hat{g}_{ab}$ ($\neq g_{ab}$) defined below. $\hat{g}_{ab}$ plays an analogous role in the laws of motion to the Newtonian effective potential $\hat{\phi}$ introduced in Sect. \ref{Sect:NewtGrav}. 

It will be seen to be most natural to choose the $\xi^a$ appearing in \eqref{BarePDef} to be generalized Killing fields associated with $\hat{g}_{ab}$. In particular, let
\begin{equation}
	\xi^{a} = \hat{g}^{ab} \Xi_{b},
\label{XiOmegaEff}
\end{equation}
where $\hat{g}^{ab}$ is the inverse of $\hat{g}_{ab}$ and $\Xi_{a}$ is a generalized Killing field constructed as described in the appendix, but with the $g_{ab}$ used there replaced by $\hat{g}_{ab}$. The worldline $\Gamma$ is chosen as a set of origins for the GKFs. This is assumed to be a timelike curve with respect to $\hat{g}_{ab}$. Similarly, the $\Sigma_s$ are assumed to be hypersurfaces formed from the set of all geodesics passing through $\gamma_s  \in \Gamma$ and orthogonal to a timelike vector field $n^a_s$ at that point (notions of geodesic, orthogonality, and timelike all being with respect to $\hat{g}_{ab}$).

%

\subsection{Initial considerations}
\label{Sect:InitConsider}

It follows from \eqref{StressCons} and \eqref{BarePDef} that $\rmd P_{\xi}(\Sigma_s)/\rmd s$ may be computed using \eqref{DeltaPTest}. In most cases of interest, the metric varies rapidly inside the body. It is therefore not useful to expand the right-hand side of \eqref{DeltaPTest} in a multipole series like \eqref{TestBodyForceExp}. We instead proceed as in the Newtonian analysis of Sect. \ref{Sect:NewtGrav}: Define a self-field using an appropriate Green function and compute the effect of this self-field on $\rmd P_\xi (\Sigma_s) /\rmd s$. The self-field should be chosen such that the force and torque exerted by it are ``ignorable'' in an appropriate sense. Subtracting the self-field away from $g_{ab}$ should also leave an ``effective metric'' $\hat{g}_{ab}$ that can be well-approximated by a Taylor series in many cases of interest. 

Quite generally, one might suppose that $g_{ab}$ can be reconstructed from an effective metric $\hat{g}_{ab}$ and a ``self-field'' $H^{ab}$ using local algebraic operations:
\begin{equation}
	g_{ab}  = G_{ab} ( \hat{g} , H ). 
	\label{MetricSplit}
\end{equation}
Assuming a relation of this sort, it follows from \eqref{DeltaPTest} that the force and torque depend on an integral involving
\begin{equation}
	T^{ab} \LieX g_{ab} = T^{ab} \big[ A_{ab}{}^{cd} ( \hat{g}, H ) \LieX \hat{g}_{cd} + B_{abcd} ( \hat{g}, H) \LieX H^{cd} \big],
	\label{ForceDensityExpand}
\end{equation}
where $A_{ab}{}^{cd}(\hat{g},H)$ and $B_{abcd}(\hat{g},H)$ are determined by $G_{ab}(\hat{g},H)$.

The first term on the right-hand side of \eqref{ForceDensityExpand} has a simple effect on $\rmd P_\xi/\rmd s$. Suppose that the $\xi^a$ are to be derived from generalized Killing fields associated with $\hat{g}_{ab}$ as described above. Comparison with \eqref{DeltaPTest} then shows that the term involving $A_{ab}{}^{cd}$ in \eqref{ForceDensityExpand} contributes forces and torques identical to those exerted on a test mass with a stress-energy tensor
\begin{equation}
	\hat{T}^{ab} := \sqrt{g/\hat{g}} T^{cd} A_{cd}{}^{ab} ( \hat{g}, H ) 
	\label{TShift}
\end{equation}
moving in a metric $\hat{g}_{ab}$. Here, $\sqrt{g/\hat{g}}$ denotes the proportionality factor relating the volume elements associated with $g_{ab}$ and $\hat{g}_{ab}$:
\begin{equation}
	\epsilon_{abcd} = \sqrt{g/\hat{g}} \hat{\epsilon}_{abcd}.
\end{equation}
In any coordinate system, this is equal to the square root of the ratio of the determinants $\det g_{\mu\nu}$ and $\det \hat{g}_{\mu\nu}$. The result is, however, independent of any coordinate choice. 

If $\LieX \hat{g}_{ab}$ can be accurately expanded in a Taylor series on a cross-section $\Sigma_s \cap W$, the first term on the right-hand side of \eqref{ForceDensityExpand} contributes forces and torques that may be expanded in a multipole series like \eqref{TestBodyForceExp}:
\begin{eqnarray}
	\fl \frac{\rmd}{\rmd s} P_\xi(\Sigma_s) = \frac{1}{2} \sum_{n=2}^{\infty} \frac{1}{n!} (I')^{c_{1} \cdots c_{n} a b}(s) 	\LieX \hat{g}_{ab, c_{1} \cdots c_{n}} (\gamma_{s}) + \frac{1}{2} \int_{\Sigma_s} T^{ab} B_{abcd} \LieX H^{cd} \rmd V.
	\label{PDotGeneralAB}
\end{eqnarray}
The $\hat{g}_{ab, c_1 \cdots c_n}$ are tensor extensions of $\hat{g}_{ab}$ as explained in Sect. \ref{Sect:TestGR}. The multipole moments $(I')^{\cdots}$ appearing here are the same as the $I^{\cdots}$ used in \eqref{TestBodyForceExp} with the replacements $g_{ab} \rightarrow \hat{g}_{ab}$ and $T^{ab} \rightarrow \hat{T}^{ab}$. These shifts may be interpreted as (finitely) ``renormalizing'' the body's quadrupole and higher multipole moments.

It is the goal of this paper to show that there is a sense in which \textit{all of} $\rmd P_\xi/\rmd s$ (or really $\rmd \hat{P}_\xi/\rmd s$ for an appropriate $\hat{P}_\xi$) can be expanded in a multipole series. Demonstrating this requires specializing further so that the second term in \eqref{PDotGeneralAB} may be simplified. At this point, $\hat{g}_{ab}$, $H^{ab}$, and $G_{ab}(\hat{g},H)$ have not yet been specified. It would be ideal if precise definitions could be provided such that
\begin{enumerate}
	\item The self-field $H^{ab}$ does not affect the force and torque in any ``essential'' way. It may shift the momenta (as is well-known to occur in electromagnetism) and the higher moments, but do nothing else. There should therefore exist some $\mathcal{E}_\xi(s)$ and some $(I'')^{c_1 \cdots c_n ab} (s)$ such that
 	\begin{equation}
		\fl \frac{1}{2} \int_{\Sigma_s} T^{ab} B_{abcd} \LieX H^{cd} \rmd V = \frac{1}{2} \sum_{n=2}^{\infty} \frac{1}{n!} (I'')^{c_{1} \cdots c_{n} a b}(s) 	\LieX \hat{g}_{ab, c_{1} \cdots c_{n}} (\gamma_{s}) - \frac{\rmd \mathcal{E}_\xi}{\rmd s} .
	\end{equation}
	The $\mathcal{E}_\xi(s)$ appearing here is to be interpreted as a ``self-momentum,'' and  must depend on properties of the system only in a finite neighborhood of $\Sigma_s$. 
	
	 \item $\hat{g}_{ab}$ satisfies the vacuum Einstein equation $\hat{R}_{ab} = \Lambda \hat{g}_{ab}$ as closely as possible in a neighborhood of $W$. Following the Newtonian analysis of Sect. \ref{Sect:NewtGrav}, one might expect that this implies ``slow variation'' of the effective metric in typical systems.
\end{enumerate}
Unfortunately, it is not clear how to realize both of these requirements exactly. We choose to implement the first precisely as stated. The second is weakened to demanding that $\hat{g}_{ab}$ approximately satisfy the vacuum Einstein equation linearized about an exact vacuum background $\bar{g}_{ab}$ if $g_{ab}$ is sufficiently close $\bar{g}_{ab}$ (although the definition of $\hat{g}_{ab}$ is non-perturbative and does not require a choice of background or that $g_{ab}$ be ``close'' to a vacuum solution in any sense). 
 
\subsection{Defining a self-field}
\label{Sect:SFDef}

Taking cues from perturbation theory in Lorenz gauge (see, e.g., \cite{PoissonRev}), choose the $G_{ab}(\hat{g},H)$ appearing in \eqref{MetricSplit} by supposing that $g_{ab}$ is equal to ``a background'' $\hat{g}_{ab}$ plus a trace-reversed perturbation $H^{ab}$:
\begin{equation}
	g_{ab} = \hat{g}_{ab} + (\hat{g}_{ac} \hat{g}_{bd} - \frac{1}{2} \hat{g}_{ab} \hat{g}_{cd}) H^{cd}.
	\label{gTogHat}
\end{equation}
Further suppose that the self-field $H^{ab}$ is related to the body's stress-energy tensor via
\begin{equation}
	H^{ab} = 4 \int_{W} \hat{G}^{aba'b'}_\rmS \hat{g}_{a'c'} \hat{g}_{b'd'} T^{c'd'} \rmd V',
	\label{HhatDef}
\end{equation}
where $\hat{G}^{aba'b'}_\rmS(x,x') = \hat{G}^{(ab)a'b'}_\rmS = \hat{G}^{ab(a'b')}_\rmS$ is an appropriately-chosen Green function associated with metric perturbations about $\hat{g}_{ab}$.

Again using Lorenz-gauge perturbation theory as a guide, let $\hat{G}^{aba'b'}_\rmS$ satisfy
\begin{eqnarray}
	\fl \hat{\Box} \hat{G}_\rmS^{aba'b'} + 2 \big\{ \hat{g}^{c(a} \hat{R}_{dcf}{}^{b)} - \frac{1}{4} \big[ \hat{g}^{ab} (\hat{R}_{df} - \Lambda \hat{g}_{df}) + ( \hat{g}^{ac} \hat{g}^{bh} \hat{R}_{ch} - \Lambda \hat{g}^{ab}) \hat{g}_{df} \big] \big\} \hat{G}_\rmS^{dfa'b'}
	\nonumber
	\\
	\qquad \qquad ~ = - 4 \pi \hat{g}^{a'c'} \hat{g}^{b'd'} \hat{g}^{(a}{}_{c'} \hat{g}^{b)}{}_{d'} \hat{\delta}(x,x') .
\label{GreenDefSimp}
\end{eqnarray}
Here, $\hat{\Box} := \hat{g}^{cd} \hat{\nabla}_c \hat{\nabla}_d$ and $\hat{g}^{a}{}_{a'}(x,x')$ may be any two-point tensor that reduces to a Kronecker-$\delta$ in the limit that its arguments coincide. For definiteness, we take $\hat{g}^{a}{}_{a'}$ to be the parallel-propagator associated with $\hat{g}_{ab}$. Given any vector $v^{a'}$ at $x'$, $g^{a}{}_{a'}(x,x') v^{a'}$ returns a vector at $x$ that is equal to $v^{a'}$ parallel-transported to $x$ using $\hat{g}_{ab}$ (along the $\hat{g}$-geodesic connecting $x$ and $x'$). If $\rmd \hat{V}$ is used to denote the volume element associated with $\hat{g}_{ab}$, the Dirac distribution $\hat{\delta}(x,x')$ satisfies
\begin{equation}
	\int_\mathcal{M} \hat{\delta}(x,x') f(x') \rmd \hat{V}' = f(x)
\end{equation}
for any test function $f(x)$ and any $x \in \mathcal{M}$. It follows from \eqref{HhatDef} and \eqref{GreenDefSimp} that $H^{ab}$ satisfies the differential equation
\begin{eqnarray}
	\fl \qquad \hat{\Box} H^{ab} + 2 \big\{ \hat{g}^{c(a} \hat{R}_{dcf}{}^{b)} - \frac{1}{4} \big[ \hat{g}^{ab} (\hat{R}_{df} - \Lambda \hat{g}_{df}) +  ( \hat{g}^{ac} \hat{g}^{bh} \hat{R}_{ch} - \Lambda \hat{g}^{ab}) \hat{g}_{df} \big] \big\} H^{df} 
	\nonumber
	\\
	\qquad \qquad \qquad \qquad \qquad ~ = - 16 \pi \sqrt{g/\hat{g}} T^{ab} .
	\label{BoxH}
\end{eqnarray}
If desired, \eqref{gTogHat} may be used obtain a similar equation for $g_{ab}$ as well (written in terms of geometric objects associated with $\hat{g}_{ab}$).

Note that Eq. \eqref{GreenDefSimp} does not define $\hat{G}^{aba'b'}_\rmS$ completely. There are many solutions to that equation (for a fixed $\hat{g}_{ab}$). A particular solution may be fixed by making the following choices \cite{PoissonRev}:
\begin{enumerate}
	\item Like the $G_\rmS$ used to define the Newtonian self-field in Sect. \ref{Sect:NewtGrav}, demand that the gravitational Green function be symmetric in its arguments:
	\begin{equation}
		\hat{G}^{aba'b'}_\rmS (x,x') = \hat{G}^{a'b'ab}_\rmS (x',x).
		 \label{GSym}
	\end{equation}
	This is essential for most of the self-force to vanish due to cancellations that can be interpreted as a manifestation of Newton's third law (as far as it can be said to hold in this context). 

	\item Also demand that $\hat{G}^{aba'b'}_\rmS (x,x')$ vanish if its arguments are timelike-separated with respect to $\hat{g}_{ab}$. This ensures that the self-field $H^{ab}$ defined by \eqref{HhatDef} does not depend on knowledge of the body in the distant past or future. 
\end{enumerate}
Enforcing both of these requirements fixes the Green function completely for a given $\hat{g}_{ab}$. The result is (at least if $\hat{R}_{ab} = \Lambda = 0$) commonly referred to as the S-type Detweiler-Whiting gravitational Green function associated with the metric $\hat{g}_{ab}$ \cite{DetWhiting, PoissonRev}.

 If $x$ and $x'$ are sufficiently close, it is known that $\hat{G}^{aba'b'}_{\rmS}(x,x')$ has the Hadamard form\footnote{Ref. \cite{PoissonRev} only derives this result in cases where $\hat{R}_{ab} = \Lambda  = 0$. The derivation is easily extended to the case considered here with no change in the conclusion.} \cite{PoissonRev}
 \begin{equation}
 	\hat{G}^{aba'b'}_{\rmS} = \frac{1}{2} [ \hat{g}^{ac} \hat{g}^{bd} \hat{g}^{(a'}{}_{c} \hat{g}^{b')}{}_{d} \hat{\Delta}^{1/2} \delta(\hat{\sigma}) - \hat{V}^{aba'b'} \Theta(\hat{\sigma}) ].
 \label{Hadamard}
 \end{equation}
 Here, $\delta$ and $\Theta$ are the Dirac and Heaviside distributions, respectively. $\hat{\sigma}(x,x') = \hat{\sigma}(x',x)$ is Synge's function computed using $\hat{g}_{ab}$, $\hat{\Delta}(x,x') = \hat{\Delta}(x',x)$ is the van Vleck determinant, and $\hat{g}^{a'}{}_{a}$ the parallel propagator. $\hat{\Delta}$ is defined to be the unique biscalar satisfying
 \begin{equation}
 	\hat{g}^{ab} \hat{\sigma}_a \hat{\nabla}_{b} \ln \hat{\Delta} = 4 - \hat{g}^{ab} \hat{\sigma}_{ab}
 \end{equation}
 and $\hat{\Delta}(x,x)=1$. This equation can be viewed as an ODE along the $\hat{g}$-geodesic connecting $x$ and $x'$. In coordinates, the solution is
 \begin{equation}
 	\hat{\Delta}(x,x') = - \frac{ \det( \hat{\nabla}_{\mu} \hat{\nabla}_{\mu'} \hat{\sigma} ) }{ \sqrt{-\hat{g}} \sqrt{ -\hat{g}'} }.
 \end{equation}
 The tail $\hat{V}^{aba'b'}(x,x') = \hat{V}^{a'b'ab}(x',x)$ appearing in \eqref{Hadamard} is a certain homogeneous solution of \eqref{GreenDefSimp}:
\begin{eqnarray}
	\hat{\Box} \hat{V}^{aba'b'} + 2 \big[ \hat{g}^{c(a} \hat{R}_{dcf}{}^{b)} - \frac{1}{4} \hat{g}^{ab} (\hat{R}_{df} - \Lambda \hat{g}_{df})
	\nonumber
	\\
	\qquad  ~ - \frac{1}{4} ( \hat{g}^{ac} \hat{g}^{bh} \hat{R}_{ch} - \Lambda \hat{g}^{ab}) \hat{g}_{df} \big] \hat{V}^{dfa'b'} = 0.
	\label{TailPDE}
\end{eqnarray}
The appropriate boundary conditions are found by integrating the ODE
\begin{eqnarray}
	\fl \hat{g}^{cd} \hat{\sigma}_c \hat{\nabla}_{d} \hat{V}^{aba'b'} + \frac{1}{2} ( \hat{g}^{cd} \hat{\sigma}_{cd} - 2) \hat{V}^{aba'b'} = \frac{1}{2} \hat{g}^{cd} \hat{\nabla}_{c}  \hat{\nabla}_{d} ( \hat{g}^{af} \hat{g}^{bh} \hat{g}^{(a'}{}_{f} \hat{g}^{b')}{}_{h} \hat{\Delta}^{1/2}  )
	\nonumber
	\\
	\qquad \qquad ~ + \big[ \hat{g}^{c(a} \hat{R}_{dcf}{}^{b)} - \frac{1}{4} ( \hat{g}^{ac} \hat{g}^{bp} \hat{R}_{cp} - \Lambda \hat{g}^{ab}) \hat{g}_{df} 
	\nonumber
	\\
	\qquad \qquad ~ -\frac{1}{4} \hat{g}^{ab} (\hat{R}_{df} - \Lambda \hat{g}_{df}) \big]  \hat{g}^{dh} \hat{g}^{fl} \hat{g}^{(a'}{}_{h} \hat{g}^{b')}{}_{l} \hat{\Delta}^{1/2} .
	\label{TailBC}
\end{eqnarray}
along all $\hat{g}$-null geodesics emanating from $x'$. $\hat{V}^{aba'b'}$ may therefore be found by solving a characteristic initial value problem. 

A Green function satisfying all of the given equations is uniquely determined if its arguments are not too widely separated (there should, in particular, exist exactly one geodesic connecting those arguments). It is important to note that the trace of $\hat{G}^{aba'b'}_\rmS (x,x')$ is always proportional to the metric. Contracting \eqref{GreenDefSimp} with $\hat{g}_{ab}$,
\begin{equation}
\hat{\Box} ( \hat{g}_{ab} \hat{G}_\rmS^{aba'b'}) + \big( 4 \Lambda - \frac{1}{2} \hat{R} \big) ( \hat{g}_{ab} \hat{G}_\rmS^{aba'b'}) = - 4 \pi \hat{g}^{a'b'} \hat{\delta} (x,x'). 
\end{equation}
It follows from this together with the imposition of appropriate boundary conditions that
\begin{equation}
\hat{g}_{ab} \hat{G}_\rmS^{aba'b'} = \hat{g}^{a'b'} \hat{G}_\rmS,
\label{GreenTrace}
\end{equation}
where $\hat{G}_\rmS(x,x')$ is the S-type Detweiler-Whiting Green function associated with a massive, nonminimally-coupled linear scalar field. $\hat{G}_\rmS$ satisfies
\begin{equation}
	\hat{\Box} \hat{G}_\rmS + \big(4 \Lambda - \frac{1}{2} \hat{R} \big) \hat{G}_\rmS = - 4 \pi \hat{\delta}(x,x')
\end{equation}
and is symmetric: $\hat{G}_\rmS(x,x') = \hat{G}_\rmS(x',x)$. This Green function also vanishes when its arguments are timelike-separated with respect to $\hat{g}_{ab}$. Explicitly,
\begin{equation}
	\hat{G}_\rmS = \frac{1}{2} \big[ \hat{\Delta}^{1/2} \delta( \hat{\sigma} ) - \hat{V} \Theta(\hat{\sigma} )\big],
\end{equation}
where
\begin{equation}
	\hat{V} = \frac{1}{4} \hat{g}_{ab} \hat{g}_{a'b'} \hat{V}^{aba'b'}.
\end{equation}

Given a physical metric $g_{ab}$ and a body with stress-energy tensor $T^{ab}$, the effective metric $\hat{g}_{ab}$ is taken to be a solution of the simultaneous equations \eqref{gTogHat}, \eqref{HhatDef}, \eqref{Hadamard}, \eqref{TailPDE}, and \eqref{TailBC}. This definition is highly implicit, and it is not clear that any solution exists (or that it is unique). Despite this, we assume that a unique solution of the given system does exist and that it has the same signature as $g_{ab}$.

One practical method for computing the effective metric is via iteration. As a first guess, suppose that $\hat{g}_{ab} \approx \hat{g}^{(0)}_{ab}$ for some metric $\hat{g}^{(0)}_{ab}$. Substituting this zeroth-order solution for $\hat{g}_{ab}$ into \eqref{gTogHat} and \eqref{HhatDef} produces the first-order approximation
\begin{equation}
	\hat{g}^{(1)}_{ab} := g_{ab} - ( \hat{g}^{(0)}_{ac} \hat{g}^{(0)}_{bd} - \frac{1}{2} \hat{g}^{(0)}_{ab} \hat{g}^{(0)}_{cd} ) H^{(0)cd} ,
	\label{DWSelfField}
\end{equation}
where 
\begin{equation}
	H^{(0)ab} := 4 \int_{W} \hat{G}^{(0) ab a'b'}_\rmS \hat{g}^{(0)}_{a'c'} \hat{g}^{(0)}_{b'd'} T^{c'd'} \rmd V'.
	\label{H0Def}
\end{equation}
$\hat{g}^{(1)}_{ab}$ may, in turn, be substituted back into \eqref{gTogHat} and \eqref{HhatDef} to find a second-order approximation for $\hat{g}_{ab}$. This process can be repeated as often as required. It is not clear that such a procedure converges uniquely for reasonable choices of $\hat{g}^{(0)}_{ab}$, although we assume that it does.

The definition of the effective metric provided here was chosen essentially as the simplest non-perturbative generalization of the Detweiler-Whiting decomposition \cite{DetWhiting, PoissonRev} typically presented for solutions of the linearized Einstein equation. It follows that $\hat{g}_{ab}$ has particularly nice properties in the linearized regime. Temporarily suppose that $g_{ab}$ is an approximate solution to Einstein's equation linearized about a background metric $\bar{g}_{ab}$ satisfying the vacuum equation $\bar{R}_{ab} = \Lambda \bar{g}_{ab}$. Define a trace-reversed metric perturbation $\mathcal{H}^{ab} = \mathcal{H}^{(ab)}$ such that
\begin{equation}
	g_{ab} = \bar{g}_{ab} + (\bar{g}_{ac} \bar{g}_{bd} - \frac{1}{2} \bar{g}_{ab} \bar{g}_{cd} ) \mathcal{H}^{cd}.
\label{TRMetricPert}
\end{equation}
If $\mathcal{H}^{ab}$ is sufficiently small\footnote{Technically, one should consider a smooth one-parameter family of metrics $g_{ab}(x;\lambda)$ and the associated perturbations $\mathcal{H}^{ab}(x;\lambda)$. For fixed $\lambda$, $g_{ab}(x;\lambda)$ is a solution to Einstein's equation with stress-energy tensor $T^{ab}(x;\lambda)$. If $g_{ab}(x;0) = \bar{g}_{ab}(x)$, it is clear that $\mathcal{H}^{ab} (x;0) = T^{ab}(x;0) = 0$. An exact solution of \eqref{LinEinstSimp} (omitting the $O(\mathcal{H}^2)$ term) is really a solution for $\partial \mathcal{H}^{ab}(x;0) / \partial \lambda$. Similarly, the $T^{ab}$ appearing in that equation should be $\partial T^{ab}(x;0)/\partial \lambda$.} and $g_{ab}$ is an exact solution to Einstein's equation with stress-energy tensor $T^{ab}$, 
\begin{eqnarray}
	\fl \bar{\Box} \mathcal{H}^{ab} + 2 \bar{g}^{c(a} \bar{R}_{dcf}{}^{b)} \mathcal{H}^{df}  + ( \bar{g}^{ab} \bar{\nabla}_d - 2 \delta_{d}^{(a} \bar{g}^{b)f} \bar{\nabla}_{f} ) \bar{\nabla}_c \mathcal{H}^{cd} =  - 16 \pi T^{ab}+ O(\mathcal{H}^2).  
\label{LinEinstSimp}
\end{eqnarray}

One may now solve for $\hat{g}_{ab}$ using the iterative method described above. It is natural in this case to use $\bar{g}_{ab}$ as a zeroth-order guess for the effective metric: $\hat{g}^{(0)}_{ab} = \bar{g}_{ab}$. Combining \eqref{DWSelfField} and \eqref{TRMetricPert} then shows that the first-order approximation for $\hat{g}_{ab}$ is given by
\begin{equation}
	\hat{g}^{(1)}_{ab} = \bar{g}_{ab} + (\bar{g}_{ac} \bar{g}_{bd} - \frac{1}{2} \bar{g}_{ab} \bar{g}_{cd} ) \mathcal{H}^{cd}_\rmR,
	\label{g1}
\end{equation}
where $\mathcal{H}^{ab}_\rmR := \mathcal{H}^{ab} - H^{(0) ab}$. Combining the first-order analog of \eqref{BoxH} with \eqref{LinEinstSimp},
\begin{eqnarray}
	\fl \qquad \qquad \bar{\Box} \mathcal{H}^{ab}_\rmR + 2 \bar{g}^{c(a} \bar{R}_{dcf}{}^{b)} \mathcal{H}^{df}_\rmR  + ( \bar{g}^{ab} \bar{\nabla}_d - 2 \delta_{d}^{(a} \bar{g}^{b)f} \bar{\nabla}_{f} ) \bar{\nabla}_c \mathcal{H}^{cd}_\rmR  =  O(\mathcal{H}^2).
\label{VacEinsLin}
\end{eqnarray}
Comparing \eqref{g1} and \eqref{VacEinsLin} with \eqref{TRMetricPert} and \eqref{LinEinstSimp} shows that $\hat{g}^{(1)}_{ab}$ is approximately a solution to the vacuum Einstein equation linearized about $\bar{g}_{ab}$. It is also expected in this linearized regime that $\hat{g}^{(1)}_{ab}$ is an excellent approximation to $\hat{g}_{ab}$. It follows that the effective metric itself is very nearly a vacuum solution if $g_{ab}$ is approximately a solution to Einstein's equation linearized about a vacuum background. Note that although the definition of the self-field was inspired by expressions for metric perturbations in Lorenz gauge, \textit{no such gauge choice is required for this conclusion}. Eq. \eqref{VacEinsLin} is the linearized Einstein equation for trace-reversed metric perturbations in any gauge. Furthermore, the equations defining $\hat{g}_{ab}$ are completely non-perturbative. They do not depend on any choice of background.
 
 \subsection{Force and torque}
 \label{Sect:FreeFall}
 
\begin{figure}
	\centering
	\includegraphics[width= .43 \linewidth]{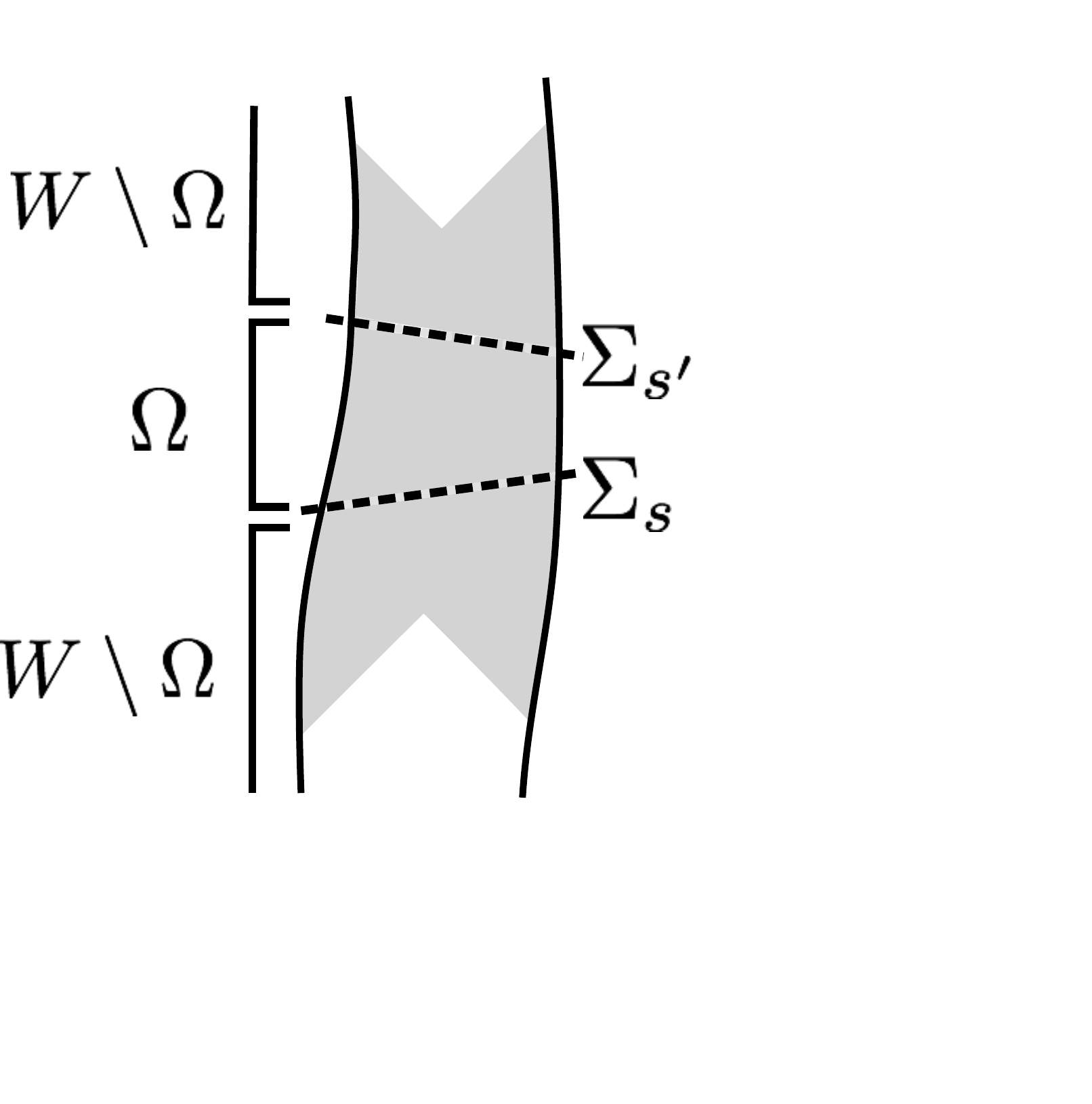}
	\vspace{-1.4 cm}
	\caption{An illustration of the spacetime regions used in \eqref{DeltaPWithF} and \eqref{FAvGen}. The shaded volume denotes the portion of the worldtube where information is required in order to compute $\mathcal{E}_\xi(\Sigma_s)$ and $\mathcal{E}_\xi(\Sigma_{s'})$.}
	\label{Fig:geometry2}
\end{figure}

Consider, once again, the momentum functional $P_\xi(\Sigma_s)$ associated with an extended mass. The (exact) difference between a component of momentum at time $s$ versus the same component at time $s' > s$ is given by \eqref{DeltaPTest}. Using this together with \eqref{gTogHat} and \eqref{HhatDef} shows that
\begin{eqnarray}
	P_\xi(\Sigma_{s'}) - P_\xi (\Sigma_{s}) &=& \frac{1}{2} \int_{\Omega(s,s')} T^{ab} (x) \LieX \hat{g}_{ab} (x) \rmd V 
	\nonumber
	\\
	&& ~+ \int_{\Omega(s,s')} \rmd V \int_W \rmd V' \mathcal{F}_\xi (x,x') ,
	\label{DeltaPWithF}
\end{eqnarray}
where
\begin{equation}
	\mathcal{F}_\xi(x,x') := 2 T^{ab} T^{a'b'} \mathcal{L}_\xi^{(x)} \Big[ (\hat{g}_{ac} \hat{g}_{bd} - \frac{1}{2} \hat{g}_{ab} \hat{g}_{cd} ) \hat{g}_{a'c'} \hat{g}_{b'd'} \hat{G}_\rmS^{cdc'd'} \Big].
\label{FDef}
\end{equation}
$\Omega(s,s') \subset W$ represents the portion of the body lying between the hypersurfaces $\Sigma_s$ and $\Sigma_{s'}$ (see Fig. \ref{Fig:geometry2}), while $\mathcal{L}_\xi^{(x)}$ denotes a ``partial Lie derivative'' that varies $x$ but not $x'$. As discussed in Sect. \ref{Sect:InitConsider}, the first term on the right-hand side of \eqref{DeltaPWithF} contributes forces and torques that are easily understood if $\hat{g}_{ab}$ varies slowly inside the body. The second term in this equation is more complicated to understand, and may be interpreted as the effect of self-interaction.

$\mathcal{F}_\xi(x,x')$ essentially represents the force exerted on matter at $x$ by matter at $x'$. For any such biscalar (even if \eqref{FDef} does not hold), note that
 \begin{eqnarray}
 	\fl \qquad \int_\Omega \rmd V \int_W \rmd V \ItF_\xi(x,x') &=& \frac{1}{2} \int_{\Omega} \rmd V 	\int_{W} \rmd V' \, [ \ItF_\xi (x,x') + \ItF_\xi (x',x) ]  \nonumber
 \\
 	&& ~ + \frac{1}{2} \int_\Omega \rmd V \int_{W \setminus \Omega} \rmd V' \, [ \ItF_\xi(x,x') - \ItF_\xi(x',x) ] .
 \label{FAvGen}
 \end{eqnarray}
if all integrals can be commuted (see Fig. \ref{Fig:geometry2}). The first line of this equation may be physically interpreted as averaging the force on matter at $x$ due to matter at $x'$ and vice-versa. In this sense, it measures the failure of Newton's 3rd law as discussed in Sect. \ref{Sect:NewtGrav}. Using \eqref{GSym}, \eqref{GreenTrace}, and \eqref{FDef},
\begin{eqnarray}
	 \fl \quad \frac{1}{2} [ \mathcal{F}_\xi (x,x') + \mathcal{F}_\xi (x',x) ] = T^{ab} T^{a'b'} \mathcal{L}_\xi ( \hat{g}_{ac} \hat{g}_{bd} \hat{g}_{a'c'} \hat{g}	_{b'd'} \hat{G}_\rmS^{cdc'd'} - \frac{1}{2} \hat{g}_{ab} \hat{g}_{a'b'} \hat{G}_\rmS ).
 \label{FAv}
\end{eqnarray}
The Lie derivative appearing here is the ordinary one acting on both $x$ and $x'$. If $\xi^{a}$ is an exact Killing vector associated with $\hat{g}_{ab}$, Eq. \eqref{FAv} vanishes exactly. In many other cases of interest, it is very small.

The second line of \eqref{FAvGen} effectively renormalizes $\ItP_{\xi}$. To see this, first note that
\begin{eqnarray}
 	\fl \qquad \quad \frac{1}{2} \int_{\Omega(s,s')} \! \rmd V \! \int_{W \setminus	\Omega (s, s') } \! \rmd V' \, [ \mathcal{F}_{\xi}	(x,x') - \mathcal{F}_{\xi}	(x',x)] = \mathcal{E}_{\xi}(\Sigma_{s}) - \mathcal{E}_{\xi}(\Sigma_{s'}),
 \label{EDef0}
 \end{eqnarray}
 where
 \begin{equation}
 	\mathcal{E}_\xi (\Sigma_s) := \frac{1}{2} \int_{\Sigma^{+}_s} \rmd V \int_{\Sigma^{-}_s} \rmd V' \, [ \mathcal{F}_\xi(x,x') - \mathcal{F}_\xi (x',x)].
 \label{EDef}
 \end{equation}
The (four-dimensional) portion of $W$ in the future of $\Sigma_s$ is denoted by $\Sigma^{+}_s$ in this equation, while the portion in its past is denoted by $\Sigma^{-}_s$. An explicit formula for $\mathcal{E}_{\xi}$ is easily obtained by combining \eqref{FDef} and \eqref{EDef}. Using the notation
 \begin{equation}
 	\hat{h}_{ab} [R] := 4 (\hat{g}_{ac} \hat{g}_{bd} - \frac{1}{2} \hat{g}_{ab} \hat{g}_{cd} ) \int_{R} \hat{g}_{a'c'} \hat{g}_{b'd'} T^{c'd'} \hat{G}^{cda'b'}_{\rmS} \rmd V'
 \label{hSDef}
 \end{equation}
 for any spacetime volume $R$,
 \begin{equation}
	 \mathcal{E}_{\xi} (\Sigma_s) = \frac{1}{4} \left( \int_{\Sigma^{+}_s} T^{ab} \LieX \hat{h}_{ab}[\Sigma^{-}_s] \rmd V -  \int_{\Sigma^{-}_s} T^{ab} \LieX \hat{h}_{ab}[\Sigma^{+}_s] \rmd V  \right).
 \label{Eexpl}	 
 \end{equation}
 Despite appearances, this depends on the behavior of the body only in a \textit{finite} four-dimensional region around $\Sigma_s \cap W$. Recalling that $\hat{G}^{aba'b'}_{\rmS} (x,x') = 0$ when $x$ and $x'$ are timelike-separated with respect to $\hat{g}_{ab}$, portions of $W$ that are timelike-separated from all of $\Sigma_s \cap W$ do not contribute to \eqref{Eexpl}. In simple cases, $\mathcal{E}_\xi(\Sigma_s)$ depends on the body's state for times of order its diameter into the past and future.  This is the shaded region in Fig. \ref{Fig:geometry2}. $\mathcal{E}_\xi$ is interpreted as a ``self-momentum'' conjugate to $\xi^a$. Further discussion may be found in Sect. \ref{Sect:EffMomenta} below.
 
An exact equation may now be written down for the force and torque acting on an extended body with a conserved stress-energy tensor. Substituting \eqref{FAvGen}-\eqref{EDef0} into \eqref{DeltaPWithF} and taking the limit $s' \rightarrow s$ shows that
\begin{eqnarray}
	 \frac{\rmd}{\rmd s} \hat{P}_\xi(\Sigma_s)  = \frac{1}{2} \int_{\Sigma_{s}} \rmd S \, T^{ab} \bigg[ \LieX \hat{g}_{ab} + 2 \int_{W} \rmd V' \, T^{a'b'}
 \nonumber
 \\
 	\qquad  \qquad ~ \times \LieX \Big(\hat{g}_{ac} \hat{g}_{bd}  \hat{g}_{a'c'} \hat{g}_{b'd'} \hat{G}^{cdc'd'}_{\rmS} - \frac{1}{2} \hat{g}_{ab} \hat{g}_{a'b'} \hat{G}_{\rmS} \Big) \bigg],
	\label{PDotExactFin}
\end{eqnarray}
where
\begin{equation}
	 \hat{\ItP}_{\xi}(\Sigma_s) := \ItP_{\xi}(\Sigma_s) + \ItE_{\xi}(\Sigma_s).
 \label{PHatDef}
\end{equation}	
Eq. \eqref{PDotExactFin} is analogous to the Newtonian expression \eqref{PDotNewt2} with two differences. First, the left-hand side involves a time derivative of $P_\xi+ \mathcal{E}_\xi$ rather than of $P_\xi$ by itself. The presence of an additional term here is interpreted as being due to the inertia of the body's self-field. That a massive object must carry a field with it as it moves affects the momentum that naturally appears in its laws of motion. The other difference between the present case and the Newtonian one is that the Lie derivative of the Green function appearing in \eqref{PDotExactFin} does not necessarily vanish. As explained in Sect. \ref{Sect:NewtGrav}, that term may be viewed as measuring the violation of Newton's third law. 

As a whole, the component of ``effective'' force or torque $\rmd \hat{\ItP}_{\xi}(\Sigma_{s}) / \rmd s$ associated with a particular vector field $\xi^a$ measures the degree to which $\hat{g}_{ab}$ is preserved by transformations in the direction $\xi^a$. To see this, note that $\hat{G}^{aba'b'}_\rmS$ is constructed in a purely geometric manner from $\hat{g}_{ab}$. This means that $\LieX \hat{G}^{aba'b'}_{\rmS}$ is a (nonlocal) linear functional in $\LieX \hat{g}_{ab}$. It therefore follows from \eqref{PDotExactFin} that $\rmd \hat{\ItP}_{\xi}(\Sigma_{s}) / \rmd s$ depends on $\xi^a$ only via terms linear in $\LieX \hat{g}_{ab}$. This dependence is confined to a finite neighborhood of $\Sigma_{s} \cap W$ sufficiently large to allow the computation of $\hat{G}_{\rmS}^{aba'b'}(x,x')$ for all null or spacelike-separated pairs $(x,x')$ with $x \in \Sigma_s \cap W$ and $x' \in W$. 

In general, \eqref{PDotExactFin} may be viewed as providing a recipe for computing $\rmd \hat{P}_\xi/\rmd s = \rmd (\ItP_{\xi} + \mathcal{E}_\xi) / \rmd s$ in terms of $\LieX \hat{g}_{ab}$. The earlier Eq. \eqref{DeltaPTest} expresses $\rmd P_\xi/\rmd s$ in terms of $\LieX g_{ab}$. These relations are mathematically equivalent. No aspect of Einstein's equation is required to transform one into the other. No properties of the $\xi^a$ are required either. All that is needed is the decomposition of $g_{ab}$ into $\hat{g}_{ab}$ and $H^{ab}$ described in Sect. \ref{Sect:SFDef}. 

The value of \eqref{PDotExactFin} over \eqref{DeltaPTest} lies in the physical expectation that $\LieX \hat{g}_{ab}$ is ``better-behaved'' than $\LieX g_{ab}$. This expectation derives from the detailed definition of $\hat{g}_{ab}$, and does depend on Einstein's equation. As shown at the end of Sect. \ref{Sect:SFDef}, $\hat{g}_{ab}$ is an approximate solution to the vacuum Einstein equation at least if $g_{ab}$ is sufficiently close to a vacuum metric. One might therefore expect that in this regime, most of the details of a body's internal structure necessarily present in $g_{ab}$ do not appear in $\hat{g}_{ab}$. A finite power series expansion for $\LieX \hat{g}_{ab}$ is therefore likely to be more accurate than one for $\LieX g_{ab}$ (given reasonable choices of $\xi^a$). A finite multipole expansion of the integral in \eqref{PDotExactFin} is then more useful than a similar expansion of \eqref{DeltaPTest}.

The construction presented here is by no means claimed to be the only reasonable one. There is considerable freedom to change the definition of $\hat{g}_{ab}$ without spoiling any of the properties just described. One can, for example, produce definitions where the equivalent of \eqref{PDotExactFin} includes a three-point interaction term involving $\LieX [ \hat{G}_\rmS^{aba'b'}(x,x') \hat{G}^{c'd'a''b''}_\rmS (x',x'')]$. An appropriate modification of this sort could be extremely valuable. It might, for example, be possible to define an effective field $\tilde{g}_{ab}$ that is an \textit{exact} solution to the vacuum Einstein equation (unlike $\hat{g}_{ab}$) and such that an appropriate $\rmd \tilde{P}_\xi /\rmd s$ depends only on $\LieX \tilde{g}_{ab}$. Whether or not this is possible to accomplish in a physically reasonably way is a question that must be left for later work.

The definitions presented in this paper are ``minimal'' in that they are essentially the simplest non-perturbative generalization of the decomposition used in the Detweiler-Whiting axiom for the motion of point masses \cite{DetWhiting, PoissonRev}. They also produce equations that closely match those that are already known for non-gravitating objects with electric or scalar charge \cite{HarteEMNew, HarteScalar}. It is, unfortunately, untrue that $\hat{g}_{ab}$ is an exact solution to the vacuum Einstein equation. This does not mean that the definitions used here necessarily fail to be useful for systems where the nonlinearity of Einstein's equation cannot be ignored. The interior of, e.g., an isolated star with constant density has a metric that is (as expected) quite simple in Fermi normal coordinates based at the center of the star \cite{FermiCoords}. Even for a very compact star of this type, the metric components are accurately approximated by low-order power series. It appears very likely that a similar result holds inside similarly-simple stars in, e.g., binary systems (before tidal disruption). These comments apply to the full physical metric $g_{ab}$. The effective metric $\hat{g}_{ab}$ is likely to have an even simpler behavior than $g_{ab}$. Precise details regarding the properties of the definitions presented here are left for future work. 


\subsection{Multipole expansions}
\label{Sect:Multipole}

Given $T^{ab}$ and $\hat{g}_{ab}$, \eqref{PDotExactFin} provides a generic (and exact) prescription for the force and torque acting on an extended body. If there exists an exact Killing vector $\hat{\zeta}^a$ associated with the effective metric $\hat{g}_{ab}$,
\begin{equation}
	\Lie_{\hat{\zeta}} \hat{g}_{ab} = \Lie_{\hat{\zeta}} \hat{G}_\rmS^{aba'b'} = \Lie_{\hat{\zeta}} \hat{G}_\rmS = 0.
\end{equation}
The effective momentum $\hat{P}_{\hat{\zeta}}$ associated with $\hat{\zeta}^a$ is therefore conserved. This is analogous to the result that the bare momentum $P_\zeta$ conjugate to $\zeta^a$ is conserved if $\Lie_\zeta g_{ab} = 0$.

Generically, there are no Killing fields associated with $g_{ab}$ or $\hat{g}_{ab}$. The integral appearing in \eqref{PDotExactFin} is then rather unwieldy to evaluate directly. It would be significantly simpler if $\LieX \hat{g}_{ab}(x)$ could be expanded in a Taylor series about $\gamma_s$ that provides a good approximation for all $x$ in a sufficiently large (but finite) ball enclosing $\Sigma_s \cap W$. That such an expansion simplifies the first term on the right-hand side of \eqref{PDotExactFin} is clear from the discussion in Sect. \ref{Sect:TestGR}. That it also simplifies the second term follows from the aforementioned fact that $\LieX \hat{G}^{aba'b'}_\rmS$ is a linear functional in $\LieX \hat{g}_{ab}$. 

We now \textit{assume} that the effective metric can be accurately expanded about $\gamma_{s}$ in a Taylor series in a Riemann normal coordinate system with origin $\gamma_s$ (where the coordinate system is constructed using $\hat{g}_{ab}$). This means that, by analogy with \eqref{MetricExpand},
\begin{equation}
	\hat{g}_{a'b'} = \hat{\sigma}_{aa'} \hat{\sigma}_{bb'} \hat{g}^{ac} \hat{g}^{bd}  \sum_{n=0}^{\infty} \frac{1}{n!} \hat{X}^{f_1} \cdots \hat{X}^{f_n} \hat{g}_{cd, f_1 \cdots f_n},
	\label{gHatTaylor}
\end{equation}
where $\hat{X}^a := - \hat{g}^{ab} \hat{\sigma}_b$ and the $\hat{g}_{ab,c_{1} \cdots c_{n}}$ denote tensor extensions of $\hat{g}_{ab}$ (see Sect. \ref{Sect:TestGR}). Although the sum here is written as an infinite one, there is no guarantee that the series will converge. We instead assume that the right-hand side is an asymptotic series for the left-hand side. There then exists a certain $n$ beyond which the series should be truncated in order to obtain an optimal approximation. Despite this, we continue to display infinite upper limits and exact equality signs for simplicity.

Recall that the $\xi^a$ are taken to be GKFs with respect to $\hat{g}_{ab}$. The origins used to construct these fields lie on the worldline $\Gamma$. Using this and \eqref{gHatTaylor}, it is possible to establish the ``hatted analog'' of \eqref{LieGExpand}:
\begin{equation}
	 \fl \qquad \quad \LieX \hat{g}_{a'b'} (x') = \sum_{n=2}^{\infty} (\cdots)_{a'b' d_1 \cdots d_n}{}^{ab c_{1} \cdots c_{n}} \hat{X}^{d_1} \cdots \hat{X}^{d_n} \LieX \hat{g}_{ab,c_{1} 	\cdots c_{n}}(\gamma_{s}).
 \label{gExpand}
\end{equation} 
The omitted coefficients in this equation are complicated but calculable. Details of their properties are discussed in \cite{HarteHighMult}. That the sum cannot include contributions from $n=0,1$ follows from $\hat{g}_{ab,c}=0$ and 
\begin{equation}
	\LieX \hat{g}_{ab} |_\Gamma = \hat{\nabla}_a \LieX \hat{g}_{bc} |_{\Gamma} = 0.
\end{equation}
Substituting \eqref{gExpand} into \eqref{PDotExactFin} and integrating term by term shows that there exist tensors $\hat{I}^{c_{1} \cdots c_{n} ab} (s)$ such that
 \begin{equation}
 	\frac{\rmd}{\rmd s} \hat{\ItP}_{\xi} (\Sigma_{s})  = \frac{1}{2} \sum_{n=2}^{\infty} 
 	\frac{1}{n!} \hat{I}^{c_{1} \cdots c_{n} ab}(s) \LieX \hat{g}_{ab,c_{1} 	\cdots c_	{n}}			(\gamma_{s}).
 \label{MultipoleSelfForce}
 \end{equation}
Note that this is identical to the test mass expression \eqref{TestBodyForceExp} for $\rmd \ItP_{\xi} / \rmd s$ with the replacements
 \begin{eqnarray}
 	g_{ab} \rightarrow \hat{g}_{ab}, \qquad
 	\ItP_{\xi} \rightarrow \hat{\ItP}_{\xi} , \qquad
 	I^{c_{1} \cdots c_{n} ab} \rightarrow \hat{I}^{c_{1} \cdots c_{n} ab}.
 \end{eqnarray}
This means that if the series expansion \eqref{gHatTaylor} for $\hat{g}_{ab}$ is valid, \textit{the momenta $\hat{\ItP}_{\xi}$ associated with a self-gravitating compact body behave as though they were the momenta of a test mass with multipole moments $\hat{I}^{c_{1} \cdots c_{n} ab}$ moving in a metric $\hat{g}_{ab}$}. All direct effects of the self-field $H^{ab}$ have been absorbed into the definitions of $\hat{\ItP}_{\xi}$ and $\hat{I}^{c_{1} \cdots c_{n} ab}$. This is made more explicit in Sect. \ref{Sect:CM} below, where laws of motion are obtained for effective linear and angular momenta $\hat{p}^{a}$ and $\hat{S}^{ab}$. The arguments are essentially identical to the ones given in Sect. \ref{Sect:TestGR} for extended test masses. 
 
Note that the multipole series \eqref{MultipoleSelfForce} is useful only if an adequate approximation is obtained by truncating it at some small finite $n$. Roughly speaking, this occurs if all significant length scales associated with $\hat{g}_{ab}$ are much larger than the body's diameter. This is likely to be the case if $g_{ab}$ is very nearly a solution to Einstein's equation linearized about a vacuum background. $\hat{g}_{ab}$ is then very nearly a solution to the linearized vacuum Einstein equation. More generally, it is not clear how useful the multipole expansion may be. 

\subsection{Effective multipole moments}
\label{Sect:EffMomenta}
 
Before continuing to describe the motion of an extended body, we first note some properties of the renormalized multipole moments appearing \eqref{MultipoleSelfForce}. The effective monopole and dipole moments are contained in $\hat{\ItP}_{\xi}$. They differ from the bare moments contained in $\ItP_{\xi}$ by the $\mathcal{E}_{\xi}$ defined in \eqref{Eexpl}. Explicit formulae for the quadrupole and higher effective moments $\hat{I}^{\cdots}$ appearing in \eqref{MultipoleSelfForce} are not derived here. They may, however, be determined (at least approximately) using the techniques developed in \cite{HarteHighMult}. This requires comparing \eqref{PDotExactFin} and \eqref{MultipoleSelfForce}, and is not difficult conceptually. It does, however, require a great deal of tedious calculation. 
 
Regardless of any detailed relations between the moments and $T^{ab}$, it is clear from \eqref{MultipoleSelfForce} and the discussion of metric normal tensors in Sect. \ref{Sect:TestGR} that the effective moments have the same index symmetries as their bare counterparts $I^{\cdots}$. This means that $\hat{I}^{c_{1} \cdots c_{n} ab}$ is separately symmetric in its first $n$ and last two indices. It also satisfies \eqref{ISym} with the substitution $I^{\cdots} \rightarrow \hat{I}^{\cdots}$. Note as well that in the regime where $\hat{g}_{ab}$ is approximately a vacuum solution to Einstein's equation, certain traces of the multipole moments decouple from the laws of motion as discussed in Sect. \ref{Sect:TestGR} and in \cite{HarteHighMult}.
 
The physical meaning of the self-momentum $\mathcal{E}_\xi$ can be made considerably more transparent by specializing to the case of a stationary system that is a solution to Einstein's equation linearized off of a Minkowski metric $\bar{g}_{ab}$. Assume that $T^{ab}$ is invariant under the action of a particular time-translation vector field $\tau^a := \partial/\partial X^0$, where $X^0$ is a globally-inertial time coordinate for $\bar{g}_{ab}$. The $\xi^{a}$ (constructed using $\hat{g}_{ab}$) are then approximately equal to Minkowski Killing fields $\bar{\xi}^a$. Consider, in particular, the Killing fields independent of $X^0$ (i.e., exclude boosts). Then,
 \begin{equation}
 	\mathcal{E}_{\xi} (\Sigma_s) = - \frac{1}{4} \int_{\Sigma_s} T^{ab} \hat{h}_{ab}[W] \bar{\xi}^{c} \rmd \bar{S}_{c} 
 \end{equation}
 to lowest nontrivial order in the metric perturbation.

 Now specialize further so that the only significant component of $T^{ab}$ is proportional to $(\partial/\partial X^0) \otimes (\partial/\partial X^0)$. Also take $\Sigma_s$ to be a hypersurface of constant $X^0$ that is unbounded in every direction. Suppose as well that the only significant metric perturbation is determined by $H^{ab}$ (so there is no ``external field''). This means that $\mathcal{H}^{ab} \approx H^{ab} \approx H^{(0)ab}$ in \eqref{TRMetricPert}. The effective ``energy'' is therefore given by
 \begin{equation}
 	- \hat{P}_\tau = \int T^{00} \left( 1+ \frac{1}{8} H^{00} \right) \rmd^3 \mathbf{X}
 \end{equation} 
 to second order in the metric perturbation. This is equivalent to
 \begin{equation}
 	- P_\tau = \int [(-g) T^{00} + t^{00} ] \rmd^3 \mathbf{X}
 	\label{PNEnergy}
 \end{equation}
in the given approximation, where $t^{ab}$ is the Landau-Lifshitz tensor in the background $\bar{g}_{ab}$:
 \begin{eqnarray}
 	16\pi t^{ab} &:=& \bar{g}_{cd} \bar{g}^{fh} \bar{\nabla}_f H^{ac} \bar{\nabla}_h H^{bd} + \frac{1}{2} \bar{g}_{cd} \bar{g}^{ab} \bar{\nabla}_f H^{ch} \bar{\nabla}_h H^{df} \nonumber
 \\
 	&& ~ - 2 \bar{g}_{cd} \bar{g}^{f (a} \bar{\nabla}_h H^{b) c} \bar{\nabla}_f H^{dh} + \frac{1}{2} ( \bar{g}^{ac} \bar{g}^{bd} - \frac{1}{2} \bar{g}^{ab} \bar{g}^{cd} ) \nonumber
 \\
 	&& ~ \times ( \bar{g}_{fp} \bar{g}_{hq} - \frac{1}{2} \bar{g}_{fh} \bar{g}_{pq} ) \bar{\nabla}_c H^{fh} \bar{\nabla}_d H^{pq} .
 \label{LL}
 \end{eqnarray}
 Eq. \eqref{PNEnergy} is the usual expression for the energy used in post-Newtonian theory. 

\subsection{Center of mass motion}
\label{Sect:CM}

At this point, effective linear and angular momenta $\hat{p}^{a}(s)$ and $\hat{S}^{ab} = \hat{S}^{[ab]}(s)$ may be introduced as tensor fields on the worldline $\Gamma$ used to define the GKFs. By analogy with the test mass equation \eqref{pSFormula}, suppose that
\begin{equation}
	\hat{\ItP}_{\xi} (\Sigma_{s}) = \hat{p}^{a}(s) \Xi_{a} (\gamma_{s}) + \frac{1}{2} \hat{S}^{ab}(s) \hat{\nabla}_{a} \Xi_{b} (\gamma_{s})
	\label{pSFormula2}
\end{equation}
for all GKFs $\Xi_{a}(x)$ defined using the metric $\hat{g}_{ab}$, the worldline $\Gamma$, and the vector field $n^{a}_{s}$. The $\xi^{a}$ appearing on the left-hand side of this equation is related to $\Xi_{a}$ via \eqref{XiOmegaEff}. Furthermore, \eqref{Del2Xi} holds with the replacements $\nabla_a \rightarrow \hat{\nabla}_a$ and $R_{abc}{}^{d} \rightarrow \hat{R}_{abc}{}^{d}$. Eq. \eqref{pSFormula2} therefore implies that 
\begin{equation}
	\fl \qquad \qquad \frac{\rmd}{\rmd s} \hat{\ItP}_{\xi} = \left( \frac{ \mathrm{\hat{D}} \hat{p}^{a}}{\rmd s} - \frac{1}{2} \hat{R}_{bcd}
	{}^{a} \hat{S}^{bc} \dot{\gamma}_{s}^{d} \right) \Xi_{a} + \frac{1}{2} \left( \frac{ \mathrm{\hat{D}} \hat{S}^{ab} }{\rmd s} - 2 \hat{p}^{[a} \dot{\gamma}^{b]}_{s} \right) \hat{\nabla}_{a} \Xi_{b}.
	\label{PDotGen}
\end{equation}
This is directly analogous to the test mass relation \eqref{PDotGenTest}. Combining it with \eqref{MultipoleSelfForce} yields multipole expansions for $\hat{\rmD} \hat{p}^{a} / \rmd s$ and $\hat{\rmD} \hat{S}^{ab}/ \rmd s$ identical to \eqref{ForceDefTest}-\eqref{TorqueMultTest} after the replacements
\begin{eqnarray}
	\quad p^{a} \rightarrow \hat{p}^{a}, \quad S^{ab} \rightarrow \hat{S}^{ab}, \quad I^{c_{1} \cdots c_{n} ab} \rightarrow \hat{I}^{c_{1} \cdots c_{n} ab} \nonumber
	\\
	g^{ab} \rightarrow \hat{g}^{ab}, \quad R_{abc}{}^{d} \rightarrow \hat{R}_{abc}{}^{d}, \quad g_{ab,c_{1} \cdots c_{n}} \rightarrow \hat{g}_{ab,c_{1} \cdots c_{n}} \label{HatAssumptions1}
	\\
	\nabla_{a} \rightarrow \hat{\nabla}_{a}, \quad \rmD / \rmd s \rightarrow \hat{\rmD} / \rmd s , \quad 
	F^{a} \rightarrow \hat{F}^{a}, \quad N^{ab} \rightarrow \hat{N}^{ab}. \nonumber
\end{eqnarray}
We refer to the resulting equations as the ``hatted forms'' of their counterparts in the theory of extended test bodies.

The worldline and foliation used to construct the GKFs may now be fixed by choosing them such that
\begin{eqnarray}
	\hat{p}^{a}(s) \propto n^{a}_{s},
	\label{pProptoNFull}
	\\
	\hat{g}_{ab}(\gamma_{s}) \hat{p}^{a}(s) \hat{S}^{bc}(s) =0.
	\label{pDotSFull}
\end{eqnarray}
These are obvious generalizations of the center of mass conditions \eqref{pPropN} and \eqref{pDotS}. Unlike in that case, however, there exists no proof that \eqref{pProptoNFull} and \eqref{pDotSFull} have well-behaved solutions. We assume, however, that they do.

As in Sect. \ref{Sect:TestGR}, it is useful to choose the parameter $s$ such that $\hat{g}_{ab} \hat{p}^{a} \dot{\gamma}^{b}_{s} = - \hat{m}$, where the effective mass is defined by
\begin{equation}
	\hat{m} : = [-\hat{g}_{ab} \hat{p}^{a} \hat{p}^{b}]^{1/2}.
	\label{MassDef}
\end{equation}
We also set $\hat{g}_{ab} n^{a}_{s} n^{b}_{s} = -1$, so $\hat{p}^{a} = \hat{m} n^{a}_{s}$. The center of mass velocity is then given by \eqref{CMVelTest} with the replacements \eqref{HatAssumptions1} and $m \rightarrow \hat{m}$. Together, the hatted versions of \eqref{ForceDefTest}-\eqref{TorqueMultTest} and \eqref{CMVelTest} strongly constrain the evolution of the body's linear and angular momenta as well as its center of mass. They do not determine it completely. As in the test body case, the evolution of the quadrupole and higher moments must be specified using other methods. Additionally, the effective metric $\hat{g}_{ab}$ couples to the motion in a nontrivial way. This is the main complication in practical computations involving the gravitational self-force.

It can be useful to define a spin 1-form $\hat{S}_{a}$ via
\begin{equation}
	\hat{S}_{a} := - \frac{1}{2} \hat{\epsilon}_{abcd} n^{b}_{s} \hat{S}^{cd}.
	\label{SpinDef}
\end{equation}
The center of mass condition \eqref{pDotSFull} guarantees that all information contained in $\hat{S}^{ab}$ is also contained in $\hat{S}_{a}$. This means that \eqref{SpinDef} is invertible:
\begin{equation}
	\hat{S}^{ab} = \hat{\epsilon}^{abcd} \hat{g}_{cf} n^{f}_{s} \hat{S}_{d}.
	\label{StoS}
\end{equation}
Note that $\hat{p}^{a} \hat{S}_{a} = 0$. The hatted form of \eqref{TorqueDefTest} implies that
\begin{equation}
	\frac{ \hat{\rmD} \hat{S}_{a} }{ \rmd s } = n^{b}_{s} \left( \hat{m}^{-1} \hat{g}_{ab} \hat{S}_{c} \frac{\hat	{\rmD} \hat{p}^{c} }{ \rmd s } - \frac{1}{2} \hat{\epsilon}_{abcd}  \hat{N}^{cd} 		\right).
	\label{SpinEvolve}
\end{equation}
The $\hat{\rmD} \hat{p}^{c} /\rmd s$ appearing on the right-hand side of this equation may be eliminated using the hatted form of \eqref{ForceDefTest}. By not doing so, one may interpret the first term in \eqref{SpinEvolve} as being responsible for a kind of Thomas precession. It arises from the requirement that $\hat{p}^{a}$ and $\hat{S}_{a}$ remain orthogonal.

There is nothing that prevents the effective mass $\hat{m}$ from varying. It immediately follows from the definition \eqref{MassDef} that
\begin{equation}
	\frac{\rmd \hat{m}}{\rmd s} = - \hat{g}_{ab} n^{a}_{s} \frac{ \hat{\rmD} \hat{p}^{b} }{\rmd s}.
\end{equation}
Substituting the hatted form of \eqref{ForceDefTest} into this equation and simplifying with the hatted form of \eqref{CMVelTest} leads to a (large) equation that does not explicitly involve $\dot{\gamma}^{a}_{s}$. Another useful form is \cite{Dix70a}
\begin{equation}
	\frac{\rmd \hat{m} }{\rmd s} = \hat{g}_{ab} \left( - \dot{\gamma}^{a}_{s} \hat{F}^{b} + \hat{m}^{-1} 	 n^{a}_{s} \hat{N}^{bc} \hat{g}_{cd} \frac{ \hat{\rmD} \hat{p}^{d} }{\rmd s} \right),
	\label{MassEvolve}
\end{equation}
which follows from \eqref{pDotSFull} as well as the hatted versions of \eqref{ForceDefTest} and \eqref{TorqueDefTest}. Additional manipulations to the right-hand side of this equation may be used to bring it into a form involving total $s$-derivatives and ``induction terms'' that depend on derivatives of the moments in a certain non-rotating reference frame \cite{Dix79, Dix70a}. Regardless, multipole expansions for $\hat{F}^a$ and $\hat{N}^{ab}$ start at quadrupole order. Multipole expansions for $\rmd \hat{m} / \rmd s$ therefore start at quadrupole order as well.

\section{Motion of a small mass: monopole and dipole approximations}
\label{Sect:Dipole}

As a simple application of the laws of motion just derived, consider the motion of a small body around a much larger one. It is possible to adopt precise approximation schemes for such problems and proceed rigorously (see, e.g., \cite{GrallaWald, PoundSF}). We instead present what is essentially a plausibility argument. This is straightforward to improve, although the details are not particularly interesting.

First consider truncating all expressions for the motion at dipole order. This corresponds to ignoring the quadrupole and higher moments $\hat{I}^{\cdots}$. Roughly speaking, this is a good approximation if all significant length scales associated with $\hat{g}_{ab}$ are sufficiently large compared with the body's own size. If $\hat{g}_{ab}$ is nearly a vacuum metric (as occurs in the regime of linearized gravity described in Sect. \ref{Sect:SFDef}) there is also a sense in which ignoring the higher moments corresponds to assuming that the body of interest is nearly spherical. Regardless of the precise reason for ignoring the higher moments, it follows from \eqref{MultipoleSelfForce} that all of the $\hat{\ItP}_{\xi}$ remain constant in this case. Applying \eqref{PDotGen} and the hatted forms of \eqref{ForceDefTest} and \eqref{TorqueDefTest} implies that the force and torque vanish: $\hat{F}^{a} = \hat{N}^{ab} = 0$. Combining this result with \eqref{MassEvolve} immediately shows that the effective mass $\hat{m}$ remains fixed in this approximation.  Furthermore, the linear and angular momenta evolve via the Papapetrou equations in the effective metric:
\begin{eqnarray}
	\frac{ \hat{\rmD} \hat{p}^{a}}{\rmd s} = \frac{1}{2} \hat{R}_{bcd}{}^{a} \hat{S}^{bc} \dot{\gamma}^	{d}_{s} ,
	\label{PapP}
	\\
	\frac{ \hat{\rmD} \hat{S}^{ab} }{\rmd s} = 2 \hat{p}^{[a} \dot{\gamma}^{b]}_{s}.	
	\label{PapS}
\end{eqnarray}
Using the hatted form of \eqref{CMVelTest}, the center of mass velocity $\dot{\gamma}^a_s$ is seen to be related to $p^a$ via
\begin{equation}
	\hat{m} \dot{\gamma}^{a}_{s} = \hat{p}^{a} + \frac{1}{2} \left( \frac{ \hat{S}^{ab} \hat{S}^{cd} {p}^{f}  	\hat{R}_{cdb}{}^{l} \hat{g}_{fl} }{\hat{m}^{2} + \frac{1}{4} \hat{S}^{bc} \hat{S}^{df} \hat{R}_{bcd}{}^{l} \hat	{g}_{fl}} \right),
	\label{PapCM}
\end{equation}
Eqs. \eqref{PapP}-\eqref{PapCM} form a coupled set of ODEs for $\hat{p}^{a}$, $\hat{S}^{ab}$, and $\gamma_{s}$. Alternatively, one may replace the evolution equation for $\hat{S}^{ab}$ with one for $\hat{S}_{a}$ using \eqref{SpinDef} and \eqref{SpinEvolve}. 

The laws of motion simplify considerably if the spin can be neglected. Note that $\hat{S}^{ab} =0$ is one solution to \eqref{PapS} and \eqref{PapCM}. It is therefore consistent to consider non-spinning bodies in the monopole-dipole approximation adopted in this section. Using \eqref{PapP} and \eqref{PapCM}, the center of mass of such a body moves on a geodesic of the effective metric:
\begin{equation}
	\frac{ \hat{\rmD} \dot{\gamma}^{a}_{s} }{ \rmd s }  = 0.
	\label{EffGeodesic}
\end{equation}
The accuracy of this equation may be estimated by evaluating, e.g., the contribution of quadrupole terms to the force and torque using the equations described above. As explained in Sect. \ref{Sect:SFDef}, $\hat{g}_{ab}$ is very nearly a vacuum solution to Einstein's equation if $g_{ab}$ is itself well-approximated by a solution to Einsteins equation (with stress-energy tensor $T^{ab}$) linearized about an exact vacuum solution. In this context, it appears reasonable that all scales associated with $\hat{g}_{ab}$ are sufficiently large compared with those of $T^{ab}$ in, e.g., fairly generic binary systems. Neglect of the higher moments is then relatively straightforward to justify. It is less clear what occurs in the highly nonlinear regime. It is likely that \eqref{EffGeodesic} remains a good approximation in many cases, although it may break down for strongly self-gravitating objects with large inhomogeneities.

Eq. \eqref{EffGeodesic} assumes that the body's spin vanishes. If $\hat{S}^{ab}/\hat{m}$ is nonzero but still small compared to length scales associated with $\hat{g}_{ab}$, Eqs \eqref{PapP}-\eqref{PapCM} may be simplified by linearizing in the spin. It then remains true that $\hat{p}^a = \hat{m} \dot{\gamma}^a_s$. Using \eqref{PapP} and \eqref{SpinEvolve},
$\hat{S}_a$ is found to be parallel-propagated along $\Gamma$ with respect to $\hat{g}_{ab}$:
\begin{equation}
	\frac{ \hat{\rmD} \hat{S}_{a} }{ \rmd s }  = 0 .
\label{EffPTrans}
\end{equation}
Substituting \eqref{StoS} into \eqref{PapP} shows that the center of mass velocity satisfies
\begin{equation}
	\hat{m} \frac{\hat{\rmD} \dot{\gamma}^a_s}{\rmd s} = \frac{1}{2} ( \hat{R}_{bcl}{}^{a} \hat{\epsilon}^{bcdf} \hat{g}_{dh} \dot{\gamma}_s^h \dot{\gamma}^l_s) \hat{S}_f
\label{PapApprox}
\end{equation}
in this case. It is unclear how consistent \eqref{EffPTrans} and \eqref{PapApprox} are over long times. They do not preserve the constraint $\hat{p}^{a} \hat{S}_{a} = 0$, which means that the spin 1-form $\hat{S}_a$ fails to remain equivalent to the spin tensor $\hat{S}^{ab}$. Alternatively, $\hat{S}^{ab}$ may be evolved instead of $\hat{S}_a$ using $\rmd \hat{S}^{ab}/\rmd s =0$. Coupling this with \eqref{PapApprox} eventually leads to violations in the center of mass condition \eqref{pDotSFull}. Such complications do not arise if Eqs. \eqref{PapP}-\eqref{PapCM} are retained in full.

The laws of motion just derived do not explicitly make any assumptions regarding the strength of a body's self-gravity (although they are more likely to be accurate when the nonlinearity of Einstein's equation can be neglected). They do not require any choice of background or gauge. More explicit expressions can be obtained by choosing a vacuum background $\bar{g}_{ab}$ and a gauge as well as assuming that $\hat{g}_{ab}$ is ``close'' to $\bar{g}_{ab}$. 

In this context, it is common to compute a body's acceleration with respect to $\bar{g}_{ab}$ rather than $\hat{g}_{ab}$. This requires writing the derivative operator $\hat{\rmD}/\rmd s$ appearing in \eqref{EffPTrans} and \eqref{PapApprox} in terms of $\bar{\rmD}/\rmd s$. For any vector $v^{a}(s)$ or covector $\omega_{a}(s)$, the appropriate transformations are
\begin{eqnarray}
	\frac{\hat{\rmD} v^{a}}{\rmd s} = \frac{ \bar{\rmD} v^{a} }{\rmd s} + \hat{C}^{a}_{bc} \dot{\gamma}^	{b}_{s} v^{c} ,
	\qquad 
	\frac{\hat{\rmD} \omega_{a}}{\rmd s} = \frac{ \bar{\rmD} \omega_{a} }{\rmd s} - \hat{C}^{c}_{ab} \dot{\gamma}^{b}_{s} \omega_{c}	,
	\label{DBar}
\end{eqnarray}
where
\begin{equation}
	\hat{C}^{a}_{bc} := \frac{1}{2} \hat{g}^{ad} ( 2 \bar{\nabla}_{(b} \hat{g}_{c)d} - \bar{\nabla}_d \hat{g}_{bc}).
\end{equation}	
To linear order in the metric perturbation $\hat{g}_{ab} - \bar{g}_{ab}$, 
\begin{equation}
	\frac{\bar{\rmD} \dot{\gamma}^a_s}{\rmd s} = \frac{ \hat{\rmD} \dot{\gamma}^{a}_{s} }{ \rmd s } + \frac{1}{2} \bar{g}^{ad} ( \bar{\nabla}_{d} 	\hat{g}_{bc} - 2 \bar{\nabla}_{b} \hat{g}_{cd} ) \dot{\gamma}^{b}_{s} \dot{\gamma}^{c}_{s} .
	\label{SelfForce}
\end{equation}
Note that in general, the background acceleration need not be orthogonal to the 4-velocity with respect to $\bar{g}_{ab}$. This is because the parameter $s$ has been chosen such that $\dot{\gamma}^{a}_{s}$ has unit norm with respect to $\hat{g}_{ab}$ (when $\hat{p}^a - \hat{m} \dot{\gamma}^a_s$ is negligible). It is more typical to normalize the 4-velocity with respect to $\bar{g}_{ab}$. Introduce a new time parameter $\bar{s}(s)$ such that $u^{a} := (\rmd s/ \rmd \bar{s}) \dot{\gamma}^{a}_{s}$ satisfies $\bar{g}_{ab} u^{a} u^{b} = -1$. Then,
\begin{equation}
	\frac{ \bar{\rmD} u^{a} }{ \rmd \bar{s} } = \left( \frac{\rmd s}{\rmd \bar{s}} \right)^2 \left( \delta^a_b - \frac{\dot{\gamma}_s^a \dot{\gamma}^c_s \bar{g}_{bc} }{ \dot{\gamma}^d_s \dot{\gamma}^f_s \bar{g}_{df} } \right) \frac{ \bar{\rmD} \dot{\gamma}^b_s }{ \rmd s} .
\end{equation}
For sufficiently small spins and weak metric perturbations, the background acceleration is obtained by substituting \eqref{PapApprox} into this equation:
\begin{equation}
	\fl \qquad \hat{m} \frac{ \bar{\rmD} u^{a} }{ \rmd \bar{s} } = \frac{1}{2} \hat{m} (\bar{g}^{ad} + u^{a} u^{d} ) ( \bar{\nabla}_{d} 		\hat{g}_{bc} - 2 \bar{\nabla}_{b} \hat{g}_{cd} ) u^{b} u^{c} + \frac{1}{2} \bar{R}_{bcl}{}^{a} \bar{\epsilon}^{bcdf} \bar{g}_{dh} u^h u^l \hat{S}_f.
	\label{SelfForceProj}
\end{equation}
 The first group of terms on the right-hand side of this expression is what is typically referred to as the gravitational self-force in the literature \cite{PoissonRev, GrallaWald, PoundSF, MST, QuinnWald}. Note, however, that this terminology differs from the non-perturbative notion of self-force adopted elsewhere in this paper. The second term on the right-hand side of \eqref{SelfForceProj} is an approximation for the ordinary Papapetrou force in the background spacetime.
 
There are more ``bulk parameters'' describing an extended body than only its center of mass position. Continuing to define quantities in terms of $\bar{g}_{ab}$, one might be interested in a ``background mass''  $\bar{m} := [-\bar{g}_{ab} \hat{p}^{a} \hat{p}^{b}]^{1/2}$. While it has already been noted that the $\hat{m}$ defined by \eqref{MassDef} remains constant in the current approximation, the same is not true of $\bar{m}$. This evolves via
\begin{equation}
	\frac{ \rmd }{ \rmd \bar{s}} \left\{  \bar{m} \left[ 1- \frac{1}{2} (\hat{g}_{ab} - \bar{g}_{ab}) u^{a} u^{b} \right] \right\} = 0
	\label{MassEvolveDipole}
\end{equation} 
to first order in the metric perturbation.

The spin evolution equation \eqref{EffPTrans} may be rewritten in terms of the background metric in the same way that Eq. \eqref{PapApprox} for the center of mass acceleration is transformed into \eqref{SelfForceProj}:
\begin{equation}
	\frac{ \bar{\rmD} \hat{S}_{a} }{ \rmd \bar{s} }  = - \frac{1}{2} \bar{g}^{cd} ( \bar{\nabla}_{d} \hat{g}_{ab} - 2 \bar{\nabla}_{(a} \hat{g}_{b)d} ) u^{b} \hat{S}_{c}.
	\label{SelfTorque}
\end{equation}
Generalizing the terminology typically applied to \eqref{SelfForceProj}, the right-hand side of \eqref{SelfTorque} might be said to represent a ``self-torque.'' This usage is, however, intrinsically perturbative. It differs from the notion of self-torque used in other parts of this paper. 

Eqs. \eqref{SelfForceProj} and \eqref{SelfTorque} describe the center of mass and spin evolution of a body with small spin and negligible higher moments if $\hat{g}_{ab}$ lies sufficiently close to a background metric $\bar{g}_{ab}$. More can be said by assuming that $g_{ab}$ also lies near $\bar{g}_{ab}$ and by adopting Lorenz gauge. This gauge choice corresponds to ensuring that the $\mathcal{H}^{ab}$ appearing in \eqref{TRMetricPert} satisfies
\begin{equation}
	\bar{\nabla}_a \mathcal{H}^{ab} = 0.
\end{equation}
Using this in \eqref{LinEinstSimp} shows that to linear order in the metric perturbation,
\begin{equation}
	\bar{\Box} \mathcal{H}^{ab} + 2 \bar{g}^{c(a} \bar{R}_{dcf}{}^{b)} \mathcal{H}^{df} = - 16 \pi T^{ab}.
	\label{LinEinstLorenz}
\end{equation}

The physical metric may be found by specifying data on an initial Cauchy surface $\Sigma_{s_0}$ and using the retarded Green function $\bar{G}^{aba'b'}_{\mathrm{ret}}(x,x')$ to evolve this data into the future (where $\bar{G}^{aba'b'}_{\mathrm{ret}}$ satisfies \eqref{GreenDefSimp} with all hats changed to bars). Then
\begin{equation}
	\mathcal{H}^{ab} = 4 \int_{\Sigma^+_{s_0}} \bar{g}_{a'c'} \bar{g}_{b'd'} \bar{G}^{aba'b'}_{\mathrm{ret}} T^{c'd'} \rmd \bar{V}' + \mathcal{H}_0^{ab} ,
\end{equation}
where $\mathcal{H}^{ab}_0$ is some homogeneous solution of \eqref{LinEinstLorenz} and $\Sigma^{+}_{s_0}$ is the portion of $W$ in the future of $\Sigma_{s_{0}}$.

Far outside of $W$, a body with slow internal dynamics produces a metric perturbation nearly indistinguishable from a retarded solution to the linearized Einstein equation with a point particle source (see, e.g., \cite{PoissonRev, GrallaWald, PoundSF}). To a first approximation, this ``effective particle'' can be taken to have the worldline $\Gamma$, mass $\hat{m}$, and no higher moments. This means that
\begin{equation}
	\fl \mathcal{H}^{ab}(x) \rightarrow 4 \hat{m} \int_{s_{0}}^{\infty} \bar{G}^{aba'b'}_{\mathrm{ret}}(x,\gamma_{s'}) \bar{g}_{a'c'}(\gamma_{s'}) \bar{g}_{b'd'}(\gamma_{s'}) u^{c'}(s') u^{d'}(s') \rmd s' + \mathcal{H}^{ab}_0(x)
	\label{LargeRH}
\end{equation}
at large distances. Now, the motion is determined by the difference field $\mathcal{H}^{ab}_\rmR$ appearing in \eqref{g1}. This is a homogeneous solution of the linearized Einstein equation in Lorenz gauge, so it may be written in Kirchhoff form:
\begin{equation}
	\fl \qquad \quad \mathcal{H}_\rmR^{ab} = \frac{1}{4\pi} \oint_{T} \bar{g}_{a'd'} \bar{g}_{b'f'} \bar{g}^{c'h'} 
	( \bar{G}^{aba'b'}_\rmS \bar{\nabla}_{c'} \mathcal{H}^{d'f'} - \bar{\nabla}_{c'} \bar{G}^{aba'b'}_\rmS \mathcal{H}^{d'f'}) 	\rmd \bar{S}_{h'} .
\end{equation}
Here $T$ is any closed hypersurface enclosing the point $x$ at which the left-hand side is evaluated.

Allowing $T$ to be a very large tube with timelike sides surrounding $W$, use of \eqref{LargeRH} shows that
\begin{equation}
	\mathcal{H}_\rmR^{ab} = 4 \hat{m} \int_{s_{0}}^{\infty} \bar{G}^{aba'b'}_{\mathrm{R}} \bar{g}_{a'c'} \bar{g}_{b'd'}u^{c'} u^{d'} \rmd s' + \mathcal{H}_0^{ab},
	\label{HR}
\end{equation}
where $\bar{G}^{aba'b'}_{\mathrm{R}} := \bar{G}^{aba'b'}_{\mathrm{ret}} - \bar{G}^{aba'b'}_{\rmS}$ is typically referred to as the R-type Detweiler-Whiting Green function (even though this satisfies a homogeneous wave equation, and is therefore not strictly a Green function). The conclusion of this argument is that for a sufficiently small particle with slow internal dynamics, the effective metric inside $W$ is essentially that of a point particle. A similar statement cannot be made for the retarded field.

The Hadamard form \eqref{Hadamard} for $\bar{G}^{aba'b'}_{\rmS}$ (and the equivalent for $\bar{G}^{aba'b'}_{\mathrm{ret}}$) may now be used to compute $\mathcal{H}_\rmR^{ab}$ explicitly. As can be seen from \eqref{SelfForceProj}, we need only the first derivative of this field on $\Gamma$. This has already been computed in, e.g., \cite{PoissonRev} for the case where $\bar{R}_{ab} = 0$ and $\bar{\rmD} u^{a} / \rmd \bar{s} = 0$. We shall continue to assume that the acceleration is zero, as any terms involving it will be negligibly small. We do, however, generalize the Ricci tensor to be $\Lambda \bar{g}_{ab}$. Then
\begin{eqnarray}
	\bar{\nabla}_{c} \mathcal{H}_\rmR^{ab}(\gamma_{s}) &=& 4 \hat{m} \left[ \left( \bar{R}_{cdf}{}^{(a} u^{b)} - u_{c} \bar{R}_{d}{}^{(a}{}_{f}{}^{b)} \right) u^{d} u^{f} - \frac{1}{3} \Lambda u_{c} u^{a} u^{b} \right]
	\nonumber
	\\
	&& ~ + H_{c}{}^{ab} + \bar{\nabla}_{c} \mathcal{H}^{ab}_0,
	\label{HRPoint}
\end{eqnarray}
where
\begin{equation}
	H_{c}{}^{ab}:= 4 \hat{m} \lim_{\epsilon \rightarrow 0} \int_{s_{0}}^{s-\epsilon} \bar{\nabla}_{c} \bar{G}^{aba'b'}_		{\mathrm{ret}} u_{a'} u_{b'} \rmd s'.
\end{equation}
For simplicity, indices in these equations have been raised and lowered with the background metric.  Also note that the limiting process used to define $H^{ab}{}_{c}$ avoids the singularity in the retarded Green function. 

Now suppose for simplicity that $\bar{\nabla}_{c} \mathcal{H}^{bc}_0$ is negligible, as can be arranged if linear perturbation theory may be trusted sufficiently far in the past. Substituting \eqref{HRPoint} into \eqref{SelfForceProj} then yields
\begin{equation}
	\fl \qquad \quad \frac{\bar{\rmD} u^{a} }{ \rmd \bar{s}} = \frac{1}{2} (\bar{g}^{ad} + u^{a} u^{d}) u^{b} u^{c} (	H_{dbc}  - 2 H_{bcd}  )  + \frac{1}{2} \bar{R}_{bcl}{}^{a} \bar{\epsilon}^{bcdf} \bar{g}_{dh} u^h u^l \hat{S}_f.
	 \label{MiSaTaQuWa}
\end{equation}
Excluding the spin term, this is the MiSaTaQuWa equation as it is usually written (at least if $s_{0} \rightarrow - \infty$) \cite{PoissonRev, GrallaWald, PoundSF, MST, QuinnWald}. Ref. \cite{GrallaWald} (which includes the spin term) refers to \eqref{MiSaTaQuWa} as the ``self-consistent'' equation of motion. This is to distinguish it from the acceleration of a deviation vector for the center of mass worldline away from a background worldline that is geodesic with respect to the background. Such constructions have not been used here.

Lastly, note that an analog of the MiSaTaQuWa equation for the spin evolution is easily obtained by substituting \eqref{HRPoint} into \eqref{SelfTorque}:
\begin{equation}
	\frac{ \bar{\rmD} \hat{S}_{a} }{\rmd \bar{s}} = - 2 \hat{m} u^{b} u^{c} \bar{R}_{abc}{}^{d} \hat{S}_{d} - \frac{1}{2} u^{b} \hat{S}^{c} ( H_{cab} - 2 H_{(ab)c} )  .
	\label{SelfTorqueMiSa}
\end{equation}

\section{Discussion}
\label{Sect:Summary}

This paper non-perturbatively defines linear and angular momenta $\hat{p}^a$ and $\hat{S}^{ab}$ adapted to the study of extended objects in general relativity. An effective metric $\hat{g}_{ab}$ is also constructed non-perturbatively from the physical metric $g_{ab}$ by generalizing the Detweiler-Whiting decomposition originally introduced in \cite{DetWhiting}. Using stress-energy conservation, $\hat{p}^a$ and $\hat{S}^{ab}$ are shown to evolve via the Papapetrou equations (written using $\hat{g}_{ab}$) plus corrections depending on the degree to which the effective metric fails to be symmetric with respect to a  certain set of generalized Killing fields.

Considerable simplifications arise in cases where $\hat{g}_{ab}$ can be accurately expanded in a power series throughout the body of interest (in the sense described in Sects. \ref{Sect:TestGR} and \ref{Sect:Multipole}). When this occurs, corrections to the Papapetrou equations may be expanded in series depending on a body's quadrupole and higher multipole moments. These series are formally identical to the multipole expansions provided by Dixon \cite{Dix79, Dix74, EhlRud} for extended test masses: There is a sense in which self-gravitating masses move like test bodies. The metric in which such a fictitious test body appears to fall is $\hat{g}_{ab}$ rather than $g_{ab}$. 

The conclusions of this paper may be viewed as providing a justification for the gravitational Detweiler-Whiting axiom \cite{DetWhiting} that a ``point mass'' moves on a geodesic in an effective metric produced by subtracting a certain ``S-field'' from the physical metric (if ``point mass'' is replaced by ``mass with small but finite size''). The validity of this type of statement has also been extended considerably. It applies to all multipole orders and also to the evolution of a body's angular momentum. This joins similar results that have recently been established for objects with scalar or electromagnetic charge moving in fixed background spacetimes \cite{HarteEMNew, HarteScalar}.


\subsection*{Future work}

The momenta proposed in this paper are not intended to be the final word on the subject. The multipole expansions for the force and torque derived in Sect. \ref{Sect:Multipole} require as their main assumption that $\hat{g}_{ab}$ vary slowly inside objects of interest. That this is likely to occur in ``typical'' systems is motivated in Sect. \ref{Sect:SFDef} by noting that in linearized gravity, the effective metric is very nearly a solution to the vacuum Einstein equation. This suggests (but does not strictly imply) that details of a body's internal structure necessarily present in $g_{ab}$ disappear in $\hat{g}_{ab}$. The vacuum condition is also useful in that it decouples many components of the ``complete'' multipole moments from the laws of motion. 

It is less clear what occurs in the fully nonlinear regime. In general, $\hat{g}_{ab}$ is not an exact solution to the vacuum Einstein equation. Despite this, the multipole expansion \eqref{MultipoleSelfForce} remains valid in this case as long as $\hat{g}_{ab}$ does not vary too rapidly. Indeed, a ``second-order self-force'' could be derived from this formalism for systems where the effective metric varies slowly. Slow variation appears likely inside strongly self-gravitating objects with a nearly uniform internal structure, although its precise range of validity is not clear. Additionally, many more components of an object's multipole moments are required to describe the motion if $\hat{R}_{ab} \neq \Lambda \hat{g}_{ab}$. 

It would be ideal if the definitions provided in this paper were modified so that extended objects could be shown to fall like test bodies moving in an effective metric that satisfies the vacuum Einstein equation exactly. As has been emphasized at various points, there is considerable freedom in the techniques developed here. Different choices could have been made in this work without spoiling any of its main conclusions. It is likely that some of these choices have even better properties than the ones that were taken. Perhaps a relatively simple change could be used to produce an effective metric $\tilde{g}_{ab}$ that satisfies the vacuum Einstein equation beyond the regime of linearized perturbation theory. 

\appendix

\section{Generalized Killing fields}
\label{GKFDefs}

The notion of a generalized Killing field (GKF) used in this paper was developed in \cite{HarteSyms}, where such objects were referred to as Killing-type generalized affine collineations. Their main properties  are summarized in Sect. \ref{Sect:TestGR}. For completeness, this appendix provides explicit definitions. It is heavily based on the description in  \cite{Bobbing}. 

Everything in this appendix is formulated on a spacetime $(\mathcal{M}, g_{ab})$. Note, however, that the main text discusses GKFs constructed using different metrics. Besides the geometry, a generalized Killing field $\Xi_{a}$ also requires for its construction a smooth timelike worldline $\Gamma = \{ \gamma_{s} | s \in \mathbb{R} \}$ and a future-directed timelike vector field $n^a_s \in T_{\gamma_{s}} \mathcal{M}$ defined along $\Gamma$. Note that $s$ is not required to be proper time and $n^a_{s}$ needn't lie tangent to $\Gamma$. 

A specific generalized Killing field may be fixed by choosing a time $s_{0}$ together with tensors $\mathcal{A}_a(s_{0})$ and $\mathcal{B}_{ab} = \mathcal{B}_{[ab]}(s_{0})$ at $\gamma_{s_{0}}$.  The Killing transport equations
\begin{eqnarray}
    & \frac{\rmD }{ \rmd s } \mathcal{A}_a(s) - \dot{\gamma}^b_{s} \mathcal{B}_{b a}(s) = 0
    \label{KT1}
    \\
    & \frac{\rmD }{ \rmd s } \mathcal{B}_{ab}(s)+ R_{a b c}{}^{d}(\gamma_{s}) \dot{\gamma}^c_{s} \mathcal{A}_d	(s) = 0
    \label{KT2}
\end{eqnarray}
are used to uniquely extend these tensors to all of $\Gamma$. Note that the skew symmetry of $\mathcal{B}_{ab}$ is preserved by this prescription\footnote{It is possible to use initial data for which $\mathcal{B}_{ab}$ is not skew. The vector field that eventually results generalizes a homothety or other non-Killing affine collineation \cite{HarteSyms}.}.

Now consider all pairs $(\gamma_{s},v^a)$, where $v^a \in T_{\gamma_{s}} \mathcal{M}$ is orthogonal to $n^a_{s}$. This forms a subset $T_\bot \Gamma$ of the tangent bundle $T \mathcal{M}$. For any element of $T_\bot \Gamma$, one may associate an affinely-parameterized geodesic $y(w)$ whose initial point is $y(0) = \gamma_{s}$ and whose initial tangent is $\dot{y}^a(0) = v^a$. As long as these geodesics can be extended sufficiently far, the map $(\gamma_{s},v^a) \rightarrow y(1)$ is a smooth function from $T_\bot \Gamma$ to $\mathcal{M}$. Its Jacobian is clearly invertible at (least at) every point $(\gamma_{s},0)$, so it follows from the inverse function theorem that the given map defines a diffeomorphism on some neighborhood $\mathcal{W}$ of $\Gamma$. This will be the region in which the GKFs are to be defined. It is assumed in the main text that the body whose motion is being studied always lies inside this region: $W \subset \mathcal{W}$.

We now define the GKF $\Xi_a(x)$ associated with a choice of $\mathcal{A}_a (s_{0})$ and $\mathcal{B}_{ab}(s_{0})$. The diffeomorphism just described may be used to uniquely associate $x$ with some $(\gamma_{s}, v^a) \in T_\bot \Gamma$. Use this pair to construct a geodesic $y(w)$ as before. The GKF is then be computed along $y(w)$ by solving the Jacobi (or geodesic deviation) equation
\begin{equation}
  \frac{\mathrm{D}^2 \Xi_a}{\rmd w^2} - R_{abc}{}^{d}  \dot{y}^b \dot{y}^c \Xi_d = 0, \label{Jacobi}
\end{equation}
with initial data
\begin{eqnarray}
  \Xi_a (\gamma_{s}) &= \mathcal{A}_a (s),
  \label{ADef}
  \\
  \frac{\rmD \Xi_a(\gamma_{s})}{\rmd w}  &= v^b \mathcal{B}_{ba} (s).
  \label{BDot}
\end{eqnarray}
The given equations uniquely define $\Xi_a(x)$ throughout $\mathcal W$ once $\mathcal{A}_a$ and $\mathcal{B}_{ab}$ are given at any one point on $\Gamma$. More detailed discussions may be found in \cite{HarteHighMult, HarteSyms, Bobbing}.

\ack

I am grateful for helpful comments and discussions with Robert
Wald, Samuel Gralla, and Barry Wardell. This work was partially supported by NSF grant
PHY04-56619 to the University of Chicago.


\section*{References}

\begin{thebibliography}{1}
\bibitem{Damour300} Damour T 1987 in \textit{300 Years of Gravitation} eds Hawking S W and Israel W (Cambridge: Cambridge University Press)
\bibitem{Dix79} Dixon W G 1979 in \textit{Isolated Systems in General Relativity} ed Ehlers J (Amserdam: North-Holland)
\bibitem{NewtSF} Detweiler S and Poisson E 2004 \PR D \textbf{69} 084019
\bibitem{GrallaHarteWald} Gralla S E, Harte A I, and Wald R M 2009 \PR D \textbf{80} 024031
\bibitem{HarteEMNew} Harte A I 2009 \CQG \textbf{26} 155015
\bibitem{Spohn} Spohn H 2004 \textit{Dynamics of Charged Particles and their Radiation Field} (Cambridge: Cambridge University Press)
\bibitem{LL} Landau L D and Lifshitz E M 1962 \textit{The Classical Theory of Fields} (Oxford: Pergamon Press)
\bibitem{Dirac} Dirac P A M 1938 \PRS A \textbf{167} 148
\bibitem{DetWhiting} Detweiler S and Whiting B~F 2003 \PR D
\bibitem{HarteHighMult} Harte A I 2010 \CQG \textbf{27} 135002
\bibitem{HarteScalar} Harte A I 2008 \CQG \textbf{25} 235020
\bibitem{PoissonRev} Poisson E 2004 \textit{Living Rev. Relativity} \textbf{7} 6
\bibitem{GerochTraschen} Geroch R and Traschen J 1987 \PR D \textbf{36} 1017
\bibitem{DewittBrehme} DeWitt B S and Brehme R W 1960 \textit{Ann. Phys. (NY)} \textbf{9} 220
\bibitem{LeviCivita1PN} Levi-Civita T 1937 \textit{Am. J. Math.} \textbf{59} 9
\bibitem{Damour83} Damour T 1983 in \textit{Gravitational Radiation} eds Dereulle N and Piran T (Amsterdam: North-Holland)
\bibitem{WillEquiv} Mitchell T and Will C M 2007 \PR D \textbf{75} 124025
\bibitem{GrallaWald} Gralla S E and Wald R M 2008 \CQG \textbf{25} 205009
\bibitem{PoundSF} Pound A 2010 \PR D \textbf{81} 024023
\bibitem{Mathisson} Mathisson M 1937 \textit{Acta Phys. Polon.} \textbf{6} 163
\bibitem{Papapetrou} Papapetrou A 1951 \textit{Proc. R. Acad. Soc.} A \textbf{209} 248
\bibitem{MST}	Mino Y, Sasaki M, and Tanaka T 1997 \PR D \textbf{55} 3457
\bibitem{QuinnWald} Quinn T C and Wald R M 1997 \PR D \textbf{56} 3381
\bibitem{Dix74} Dixon W G 1974 \PTRS A \textbf{277} 59
\bibitem{HarteSyms} Harte A I 2008 \CQG \textbf{25} 205008
\bibitem{Dix70a} Dixon W G 1970 \PRS A \textbf{314} 499
\bibitem{Wald} Wald R M 1984 \textit{General Relativity} (Chicago: University of Chicago Press)
\bibitem{CM} Schattner R 1979 \textit{Gen. Rel. Grav.} {\bf 10} 377; {\bf10} 395
\bibitem{HarteQuadrupole} Harte A~I 2007 \CQG \textbf{24} 5161
\bibitem{ItalianQuadrupole} Bini D, Fortini P, Geralico A and Ortolan A 2008 \CQG \textbf{25} 125007; \textbf{25} 035005
\bibitem{SchattnerStreubel1} Schattner R and Streubel M 1981 \textit{Ann. de l'Inst. H. Poincar\'{e}} \textbf{34} 117 
\bibitem{SchattnerStreubel2} Streubel M and Schattner R 1981 \textit{Ann. de l'Inst. H. Poincar\'{e}} \textbf{34} 145
\bibitem{HarteEMOld} Harte A I 2006 \PR D \textbf{73} 065006
\bibitem{Dix67} Dixon W G 1967 \textit{J. Math. Phys.} \textbf{8} 1591
\bibitem{Synge} Synge J L \textit{Relativity: The General Theory} (Amsterdam: North-Holland)
\bibitem{Friedlander} Friedlander F G 1975 \textit{The Wave Equation on a Curved Space-Time} (Cambridge: Cambridge University Press)
\bibitem{ThorneMoments} Thorne K S 1980 \textit{Rev. Mod. Phys.} \textbf{52} 299 
\bibitem{EhlRud} Ehlers J and Rudolph E 1977 \textit{Gen. Rel. Grav.} \textbf{8} 197
\bibitem{FermiCoords} Klein D and Collas A 2010 \textit{J. Math. Phys.} \textbf{51} 022501
\bibitem{Bobbing} Gralla S E, Harte A I, and Wald R M 2010 \PR D \textbf{81} 104012
\end{thebibliography}
\end{document}